\documentclass[extra,mreferee]{gji}
\journal{}
\usepackage{timet,color}
\usepackage[urlcolor=blue,citecolor=black,linkcolor=black]{hyperref}

\usepackage{amsthm}
\usepackage{amsmath}
\usepackage{amssymb}
\usepackage{graphicx}
\usepackage{subcaption}
\usepackage{etoolbox}

\graphicspath{{Figures/}}

\newtheoremstyle{lemma}
  {10pt} {10pt} {\itshape} {} {\bfseries} {.} {.5em} {}
\theoremstyle{lemma}

\usepackage{aliascnt}
\usepackage{cleveref}
\usepackage{listings}
\usepackage{xcolor}
\usepackage{booktabs}

\definecolor{codegray}{gray}{0.95}
\definecolor{commentgreen}{rgb}{0,0.6,0}
\definecolor{keywordblue}{rgb}{0.2,0.2,0.8}
\definecolor{stringred}{rgb}{0.6,0.1,0.1}

\lstdefinestyle{mypython}{
    language=Python,
    backgroundcolor=\color{codegray},
    basicstyle=\ttfamily\footnotesize\linespread{1.1}, 
    keywordstyle=\color{keywordblue}\bfseries,
    commentstyle=\color{commentgreen}\itshape,
    stringstyle=\color{stringred},
    showstringspaces=false,
    breaklines=true,
    breakatwhitespace=true,
    keepspaces=true,
    columns=flexible,
    frame=single,
    tabsize=4,
    captionpos=b,
    numbers=none,
    xleftmargin=1em,
    xrightmargin=1em,
}

\title[EEPAS Software]
  {The EEPAS Model Revisited: Statistical Formalism and a High-Performance, Reproducible Open-Source Framework}
\author[Chung et. al.]
 {Szu-Chi Chung$^1$, 
  Chien-Hong Cho$^1$\thanks{Corresponding author: chcho@math.nsysu.edu.tw} and 
  Strong Wen$^2$ \\ 
  $^1$ Department of Applied Mathematics, National Sun Yat-sen University, Kaohsiung, Taiwan.\\
  $^2$ Department of Earth and Environmental Sciences, National Chung Cheng University, Chia-Yi, Taiwan.
 }
\pagerange{\pageref{firstpage}--\pageref{lastpage}}

\makeatletter
\patchcmd{\eqsecnum}
  {\normalsize} 
  {}            
  {\message{Patched \string\eqsecnum (gji.cls)}} 
  {\message{ERROR: Failed to patch \string\eqsecnum}} 

\patchcmd{\eqsubsecnum}
  {\normalsize}
  {}
  {\message{Patched \string\eqsubsecnum (gji.cls)}}
  {\message{ERROR: Failed to patch \string\eqsubsecnum}}

\apptocmd{\appendix}
  {%
    \renewcommand\theequation{\hbox{\Alph{section}.\arabic{equation}}}%
    \message{Patched \string\appendix's \string\theequation (gji.cls)}%
  }
  {}
  {\message{ERROR: Failed to patch \string\appendix}} 
\makeatother

\begin{document}

\label{firstpage}

\maketitle

\begin{summary}

While short-term models such as the Short-Term Earthquake Probability (STEP) and Epidemic-Type Aftershock Sequence (ETAS) are well established and supported by open-source software, medium- to long-term models, notably the Every Earthquake a Precursor According to Scale (EEPAS) and Proximity to Past Earthquakes (PPE), remain under-documented and largely inaccessible. Despite outperforming time-invariant models in regional studies, their mathematical foundations are often insufficiently formalized. This study addresses these gaps by formally deriving the EEPAS and PPE models within the framework of inhomogeneous Poisson point processes and clarifying the connection between empirical $\Psi$-scaling regressions and likelihood-based inference. We introduce a fully automated, open-source Python implementation of EEPAS that combines analytical modeling with \texttt{Numba} JIT acceleration, \texttt{NumPy} vectorization, and \texttt{joblib} parallelization, all configured via modular JSON files for usability and reproducibility. Integration with \texttt{pyCSEP} enables standardized evaluation and comparison. When applied to the Italy HORUS dataset, our system reproduces published results within one hour using identical initialization settings. It also provides a comprehensive pipeline from raw catalog to parameter estimation, achieving improved log-likelihoods and passing strict consistency tests without manual $\Psi$ identification. We position our framework as part of a growing open-source ecosystem for seismological research that spans the full workflow from data acquisition to forecast evaluation. Our framework fills a key gap in this ecosystem by providing robust tools for medium- to long-term statistical modeling of earthquake catalogs, which is an essential but underserved component in probabilistic seismic forecasting.

\end{summary}


\section{Introduction}

Earthquake forecasting refers to the probabilistic assessment of earthquake occurrence within a defined time, region, and magnitude range. Earthquakes arise from processes that exhibit both deterministic and stochastic features, with their most predictable aspects being spatial and temporal clustering and frequency–magnitude distributions. Developing reliable forecasting models and understanding earthquake generation are central goals of statistical seismology, requiring the integration of physical and statistical insights. Core empirical laws, such as the Omori–Utsu law for aftershock decay and the Gutenberg–Richter law for magnitude distributions, were statistically derived well before their physical bases were understood. In statistical seismology, earthquake forecasting models are typically categorized as short-term, medium-term, or long-term. Well-known short-term models include the Short-Term Earthquake Probability (STEP) and Epidemic-Type Aftershock Sequence (ETAS) models. Both are supported by publicly available and well-maintained open-source implementations \cite{jalilian2019etas,mizrahi2021effect,lombardi2017seda,field2003opensha}. In contrast, medium to long-term models, such as the Every Earthquake a Precursor According to Scale (EEPAS) \cite{RE} and Proximity to Past Earthquakes (PPE) \cite{JK} models, remain largely inaccessible. To the best of our knowledge, no actively maintained open-source implementations of these models currently exist. The only available EEPAS implementation uses proprietary MATLAB software \cite{2023ER}, which is also difficult to use and extend due to hard-coded paths, undocumented variables, and inefficient computation. These issues hinder reproducibility and large-scale experimentation.

Despite these limitations, EEPAS and PPE are well-established medium to long-term forecasting models based purely on seismicity. They have been successfully applied to several regional catalogs and consistently outperform time-invariant models in forecasting major earthquakes \cite{2023ER,RE,rhoades2011,rhoades202220}. However, their mathematical foundations are insufficiently documented. First, the EEPAS model is built on empirical linear relationships observed in seismic data, known as the $\Psi$ phenomenon. Yet, this phenomenon has only been superficially analyzed. Early studies relied on manual identification and lacked an objective procedure. Although a recent “rectangular algorithm” was proposed to automate $\Psi$ detection \cite{psi}, it has not been validated within the EEPAS end-to-end forecasting framework. The associated code is incomplete, lacking support for resolving multiple $\Psi$ events, visualization routines, and derivation of initialization values for EEPAS. Second, existing studies rarely provide a formal explanation for how deterministic $\Psi$ relationships are transformed into the probabilistic distributions used in the model’s likelihood function. Third, key components of the EEPAS log-likelihood, specifically the incompleteness correction factor $\Delta(m)$ and the normalization factor $\eta(m)$, are often presented without formal proofs or transparent derivations. Finally, while Maximum Likelihood Estimation (MLE) of PPE parameters is known to yield models in which the total expected number of events matches the observed count, the theoretical justification for this property is seldom discussed in the literature.

To address these gaps, this study clarifies the mathematical foundations of the PPE and EEPAS models. We derive analytical expressions for both $\Delta(m)$ and $\eta(m)$, revealing how the model accounts for missing data and scales precursor productivity. We formally derive the complete EEPAS log-likelihood, grounding it in the framework of inhomogeneous Poisson point processes. Furthermore, we provide a rigorous rationale linking the empirical $\Psi$ relationships to their probabilistic formulations, thereby strengthening the statistical foundation of the model. We also present a mathematical proof demonstrating that, under specific conditions, the joint MLE procedure inherently yields a model whose rate function integrates to the total number of observed events. 


Building on recent theoretical developments, we have created a fully automated pipeline that transforms raw earthquake catalogs into initial EEPAS parameter estimates. The EEPAS model has been entirely re-engineered in a modern, high-performance Python framework that adheres to the original methodology while significantly improving computational efficiency. Where possible, numerical integrations have been replaced with analytical solutions to reduce computational overhead. Core routines have been optimized through vectorization using NumPy \cite{harris2020array} and Numba’s just-in-time (JIT) compilation \cite{lam2015numba}, while parallel processing of independent tasks, such as cell-based integrations, is implemented using \texttt{joblib} \cite{joblib-software}. A centralized JSON configuration file manages all parameters, file paths, and runtime settings, enhancing clarity, reproducibility, and extensibility.  The new implementation integrates seamlessly with the broader forecasting workflow. It supports upstream preprocessing with SeismoStats \cite{mirwald2025seismostats} for initial seismic parameter estimation and downstream evaluation using tools such as \texttt{pyCSEP} \cite{savran2022pycsep,graham2024new}, allowing for rigorous statistical testing and model comparison. The codebase is thoroughly documented, with clear explanations provided for each module and configuration option. This framework fills a critical gap in the open-source seismological ecosystem by offering robust support for medium to long-term statistical earthquake modeling, an essential yet often underrepresented component of probabilistic seismic hazard assessment.

\section{Mathematical foundation}


\begin{figure}
\centering
\includegraphics[width=\textwidth]{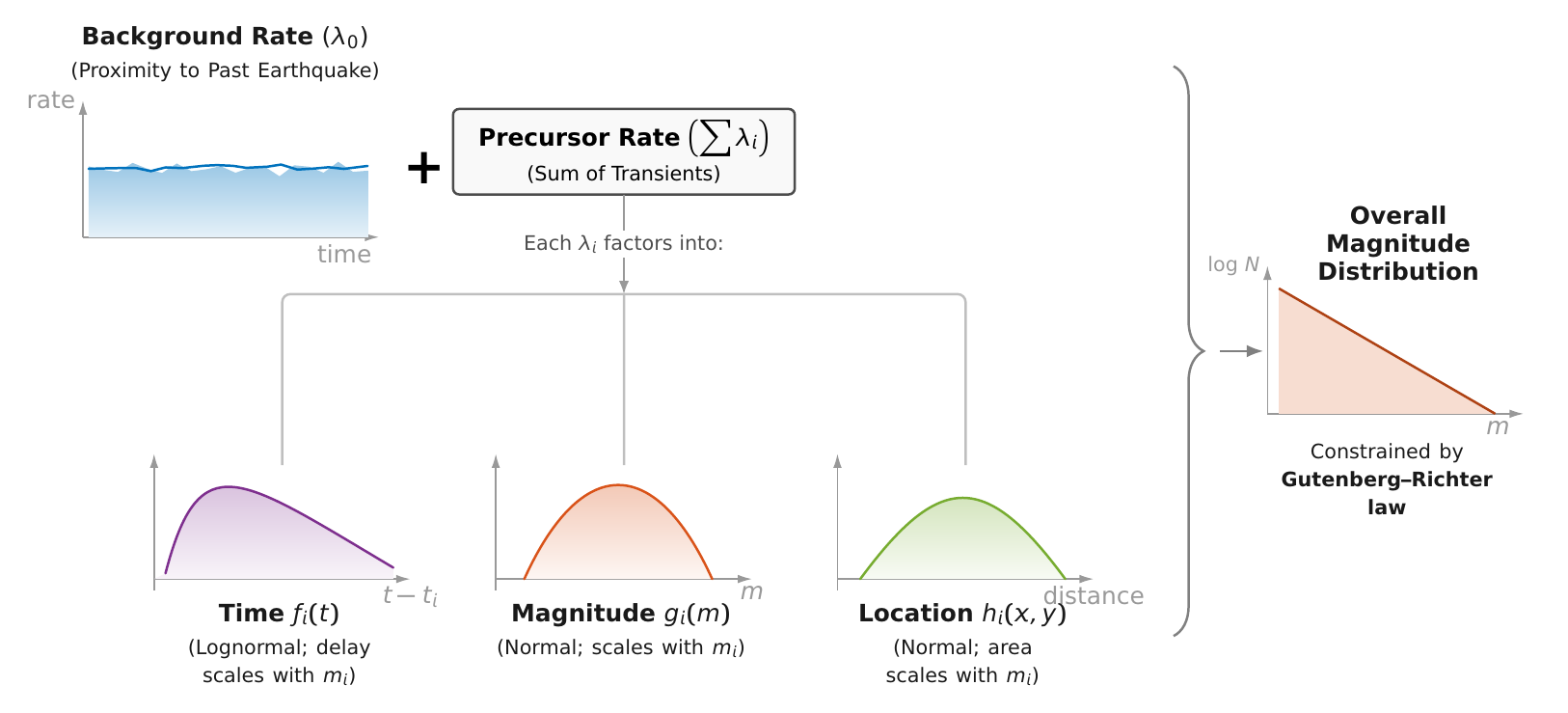}
\caption{Schematic illustration of the EEPAS model. The total earthquake rate consists of a baseline background rate and a sum of precursor rates. Each precursor rate is composed of three factorized density components (time, magnitude, and space). All earthquakes, whether background or precursor events, are constrained to follow the same magnitude distribution as defined by the Gutenberg–Richter (GR) law. 
}
\label{fig:model}
\end{figure}



The EEPAS model assumes that every earthquake is a potential precursor, with each precursor contributing to the overall rate density. Let the $i$-th earthquake (indexed by its occurrence time), observed in a data collection region $S$, occur at time $t_i \geq t_0$, with magnitude $m_i \geq m_0$ and location $(x_i, y_i)$. Each such event contributes an instantaneous increment $\lambda_i(t,m,x,y)$ to the rate density of future seismicity in its neighborhood, defined as
\begin{equation}\label{li}
    \lambda_i(t,m,x,y) = w_i f_i(t) g_i(m) h_i(x,y),
\end{equation}
where $w_i$ denotes a weighting factor determined by the aftershock model proposed in \cite{Rhoades}. Here, $t_0$ is the time from which the earthquake data are taken into consideration, and $m_0$ denotes the lower bound of magnitude that can be reliably recorded during the period $t \geq t_0$, based on the precision of the measuring instruments.

To account for earthquakes that occur without identifiable precursors, the full EEPAS model is often combined with the PPE model and defined as
\begin{equation}\label{ltmxy}
    \lambda(t,m,x,y) = \mu \lambda_0(t,m,x,y) + \sum_{t_i \geq t_0;\, m_i \geq m_0}^{t - \mathrm{delay}} \eta(m_i) \frac{\lambda_i(t,m,x,y)}{\Delta(m)}.
\end{equation}
Here, $\lambda_0(t,m,x,y)$ represents the baseline rate density specified by the PPE model, and $\mu$ denotes the \textit{failure-to-predict rate}, defined as the proportion of earthquakes that occur without a discernible sequence of precursory events. The full integration of these two components is illustrated in \Cref{fig:model}. 

Since earthquakes below $m_0$ are excluded, their contributions to the rate are unaccounted for. To compensate for this incompleteness, each $\lambda_i(t,m,x,y)$ is scaled by a factor of $1/\Delta(m)$, where
\begin{equation}\label{eq:Delta}
\Delta(m) = \Phi\!\left(\frac{m - a_M - b_M m_0 - \sigma_M^2 \beta}{\sigma_M}\right),
\end{equation}
and $\Phi(x) = \frac{1}{\sqrt{2\pi}} \int_{-\infty}^x e^{-t^2/2}\,dt$ is the standard normal cumulative distribution function. The parameters $a_M$, $b_M$, and $\sigma_M$ define the magnitude component of EEPAS. Meanwhile, $\eta(m)$ is a magnitude-dependent scaling function defined as
\begin{equation}\label{emi}
    \eta(m) = \frac{(1 - \mu) b_M}{E(w)} \exp\!\left\{ -\beta \left[ a_M + (b_M - 1)m + \frac{\sigma_M^2 \beta}{2} \right] \right\},
\end{equation}
where $E(w)$ denotes the expected value (mean) of the weights $w_i$ for earthquakes that occurred prior to the one with weight $w$, and $\beta = b_{\mathrm{GR}} \ln 10$, with $b_{\mathrm{GR}}$ denoting the Gutenberg–Richter $b$-value.

The superscript ${}^{t - \mathrm{delay}}$ in the summation indicates that each precursor’s contribution $\lambda_i$ becomes active only after a short exclusion window (“delay”) following its origin time $t_i$. This delay is introduced to suppress short-term clustering effects (e.g., aftershocks or immediate foreshocks) that could otherwise obscure medium-term precursory patterns. The parameter $\mu \in [0,1]$ serves as a mixture weight representing the contribution from earthquakes without detectable precursors.

Accordingly, the PPE component $\lambda_0 = f_0\,g_0\,h_0$ models the expected space–time–magnitude distribution of earthquakes \emph{in the absence of medium-term precursory build-up}. It provides a baseline intensity, while the precursor component enhances the rate in regions and times where medium-term precursory scaling is evident. The scaling function $\eta(\cdot)$, normalized by $E(w)$, ensures that the integrated contribution from the precursor term averages to $(1 - \mu)$. A complete description of the EEPAS and PPE models is provided in \Cref{sec:model}.

\section{Implementation for EEPAS}\label{sec:implementation}




\subsection{PPE model fitting}

In this step, we estimate the three free parameters $a$, $d$, and $s$ by maximizing the log-likelihood function of a Poisson point process:
\begin{equation}\label{lnL0}
\begin{aligned}
\ln{\mathcal{L}(\lambda_0)} \equiv \sum_{\substack{t_j \in [t_s,t_e)\\ m_j \geq m_T\\ (x_j,y_j)\in R}} \ln{\lambda_0(t_j,m_j,x_j,y_j)}
- \int_{t_s}^{t_e}\int_{m_T}^{m_u}\iint_{R} \lambda_0(t,m,x,y)\, dA\, dm\, dt,
\end{aligned}
\end{equation}
where $t_s$ and $t_e$ are the starting and ending times of the fitting period, $m_T$ and $m_u$ are the lower and upper bounds of the magnitude range under consideration and $R \subseteq  S$ is the testing Region. Here, $S$ stands for the data collection area. All earthquakes occurring within $[t_s, t_e)$ with magnitudes in $[m_T, m_u]$ are considered in the fit. Therefore, we also call $[t_s,t_e)$ the learning period. 

From \eqref{l0}, we have:
\begin{equation}\label{l01}
\begin{aligned}
\lambda_0(t_j, m_j, x_j, y_j) = &\frac{\beta}{t_j - t_0} \exp[-\beta(m_j - m_T)] \\
&\times \sum_{\substack{t_i \in [t_0,\,t_j - \mathrm{delay})\\ m_i \geq m_T}} \left[a(m_i - m_T) \frac{1}{\pi} \left( \frac{1}{d^2 + (x_j - x_i)^2 + (y_j - y_i)^2} \right) + s \right],
\end{aligned}
\end{equation}
for all $t_j \in [t_s, t_e)$, provided $m_j \geq m_T$. The first term of \eqref{lnL0} is computed by summing over all $j$ satisfying $t_j \in [t_s, t_e)$. For the second term, we note that if an earthquake occurred before $t_s-\mathrm{delay}$ ($t_i<t_s-\mathrm{delay}$), the time integral spans $[t_s, t_e]$. If the earthquake occurred after $t_s-\mathrm{delay}$ ($t_i\geq t_s-\mathrm{delay}$) and before $t_e-\mathrm{delay}$ ($t_i<t_e-\mathrm{delay}$), the time integral spans $[t_i + \mathrm{delay}, t_e]$. Thus, the second term becomes:
\begin{equation}\label{l02}
\begin{aligned}
&\int_{t_s}^{t_e} \int_{m_T}^{m_u} \iint_{R} \lambda_0(t,m,x,y)\, dA\, dm\, dt \\
&= \sum_{\substack{t_i \in [t_0,\, t_s - \mathrm{delay})\\ m_i \geq m_T}} 
\left[ \int_{t_s}^{t_e} f_0(t)\, dt \right] 
\left[ \int_{m_T}^{m_u} g_0(m)\, dm \right] 
\left[ \iint_R h_{0i}(x,y)\, dA \right] \\
&\quad + \sum_{\substack{t_i \in [t_s - \mathrm{delay},\, t_e - \mathrm{delay})\\ m_i \geq m_T}} 
\left[ \int_{t_i + \mathrm{delay}}^{t_e} f_0(t)\, dt \right] 
\left[ \int_{m_T}^{m_u} g_0(m)\, dm \right] 
\left[ \iint_R h_{0i}(x,y)\, dA \right].
\end{aligned}
\end{equation}

\noindent Here, index $j$ refers to observed events in the fitting window $[t_s, t_e)$ with $m_j \ge m_T$, at which $\lambda_0$ is evaluated. Index $i$ refers to historical source events that contribute to the spatial proximity field via $h_{0i}$, drawn from the data collection region $S$. In \eqref{l01}, for each fixed $j$, the sum is over all $i$ with $t_i \in [t_0, t_j - \mathrm{delay})$ and $m_i \ge m_T$. In \eqref{l02}, each $i$ contributes a block whose time integral starts at $\max(t_s,\, t_i + \mathrm{delay})$. The functions $f_0$ and $g_0$ are not indexed by $i$ because they depend only on the evaluation variables $t$ and $m$. The integrals are computed as follows (noting $f_0(t) = \frac{1}{t - t_0}$ for $t > t_0$ and $g_0(m) = \beta e^{-\beta(m - m_T)}$ for $m \ge m_T$):
\begin{equation}
\begin{aligned}
\int_{t_s}^{t_e} f_0(t)\, dt &= \ln(t - t_0)\bigg|_{t_s}^{t_e} 
= \ln(t_e - t_0) - \ln(t_s - t_0), \\
\int_{t_i + \mathrm{delay}}^{t_e} f_0(t)\, dt 
&= \ln(t - t_0)\bigg|_{t_i + \mathrm{delay}}^{t_e} 
= \ln(t_e - t_0) - \ln(t_i + \mathrm{delay} - t_0), \\
\int_{m_T}^{m_u} g_0(m)\, dm 
&= \exp[-\beta(m_T - m_0)] - \exp[-\beta(m_u - m_0)].
\end{aligned}
\end{equation}

The spatial integral $\iint_R h_{0i}(x,y)\, dA$ is evaluated numerically on the finite region $R$. To find the optimal parameters $a$, $d$, and $s$, we minimize $-\ln \mathcal{L}(\lambda_0)$ using constrained Nelder–Mead optimization.

\subsection{Computing weighting coefficients \texorpdfstring{$w_i$}{wi}}
As proposed in \cite{RE}, the weighting coefficients $w_i$ are determined by a model allowing aftershocks \eqref{l'tmxy}.
To estimate the aftershock parameters, we fit $(\nu, \kappa)$ by maximizing the log-likelihood function in \eqref{lnL'} over all earthquakes with $m \ge m_T$ that occurred within region $R$ during the interval $[t_s, t_e)$, \emph{conditional on a fixed} Bath-type lower bound $\delta$.

\begin{equation}\label{lnL'}
\begin{aligned}
\ln{\mathcal{L}(\lambda')} \equiv \sum_{\substack{t_j \in [t_s,t_e)\\ m_j \ge m_T\\ (x_j,y_j) \in R}} \!\! \ln{\lambda'(t_j,m_j,x_j,y_j)} 
- \int_{t_s}^{t_e} \int_{m_T}^{m_u} \iint_{R} \lambda'(t,m,x,y)\, dA\, dm\, dt.
\end{aligned}
\end{equation}

Let $(t_j, m_j, x_j, y_j)$ denote the time, magnitude, and location of the $j$-th earthquake that occurred in the learning period $[t_s,t_e)$. For the first term in \eqref{lnL'}, since PPE model has been fitted in the previous step, $\lambda_0(t_j, m_j, x_j, y_j)$ can now be computed by \eqref{l01}. Besides, based on \eqref{l'itmxy}--\eqref{h2ixy}, the aftershock density function $\lambda_i'(t_j, m_j, x_j, y_j)$, where $t_i < t_j$, is given by:
\begin{equation*}
\begin{aligned}
\lambda_i'(t_j, m_j, x_j, y_j) ={} & \frac{p - 1}{(t_j - t_i + c)^p} \, H(m_i - \delta - m_j)\, \beta \exp\!\left[ -\beta(m_j - m_i) \right] \\
& \times \frac{1}{2\pi \sigma_U^2 10^{m_i}} \exp\!\left[ -\frac{(x_j - x_i)^2 + (y_j - y_i)^2}{2 \sigma_U^2 10^{m_i}} \right]
\end{aligned}
\end{equation*}
Here, the parameter $\delta$ is fixed at $0.7$ as was used in \cite{2023ER,RE}.  

For the second term in \eqref{lnL'}, the triple integral corresponding to the PPE background component is approximated as:
\begin{equation*}
\begin{aligned}
\int_{t_s}^{t_e} \int_{m_T}^{m_u} \iint_R \lambda_0(t, m, x, y)\, dA\, dm\, dt 
&= \int_{t_s}^{t_e} f_0(t)\, dt \int_{m_T}^{m_u} g_0(m)\, dm \iint_R h_0(x, y)\, dA \\
&= \ln \frac{t_e - t_0}{t_s - t_0} \left[ \exp\!\big[-\beta(m_T - m_0)\big] - \exp\!\big[-\beta(m_u - m_0)\big] \right] \\
& \times \iint_R h_0(x, y)\, dA
\end{aligned}
\end{equation*}
See Appendix E for the detail. On the other hand, the triple integral for each $\lambda_i'$ can be calculated as follows:
\begin{equation*}
\begin{aligned}
    \int_{t_s}^{t_e} \int_{m_T}^{m_u} \iint_R \nu\, \lambda_i'(t, m, x, y)\, dA\, dm\, dt 
    & = \int_{t_s}^{t_e} f_i'(t)\, dt \int_{m_T}^{m_u} g_i'(m)\, dm \iint_R h_i'(x, y)\, dA \\
    & = I_t(t_s, t_e \mid t_i)I_m(m_T, m_u \mid m_i, \delta)\iint_R h_i'(x, y),
\end{aligned}
\end{equation*}
where
\begin{align}    
I_t(t_s, t_e \mid t_i) 
= \int_{\max\{t_s, t_i\}}^{t_e} \frac{p - 1}{(t - t_i + c)^p}\, dt \notag 
= \left(\max\{t_s, t_i\} - t_i + c \right)^{1 - p} - (t_e - t_i + c)^{1 - p},
\end{align}
and
\begin{align}
I_m(m_T, m_u \mid m_i, \delta) 
= \int_{m_T}^{\min\{m_u, m_i - \delta\}} \beta\, e^{-\beta(m - m_i)}\, dm \notag 
= e^{-\beta(m_T - m_i)} - e^{-\beta(\min\{m_u, m_i - \delta\} - m_i)}.
\label{Imj}
\end{align}
For the integral of $h_i'$, we compute the exact integral of the bivariate Gaussian over each rectangular tile $[X_1, X_2] \times [Y_1, Y_2]$ (in km) covering $R$. With $\sigma_i = \sigma_U\, 10^{m_i/2}$, the integral is:
\begin{equation*}
\begin{aligned}
\iint_{[X_1, X_2] \times [Y_1, Y_2]} h'_i(x, y)\, dA 
&= \iint_{[X_1, X_2] \times [Y_1, Y_2]} \frac{1}{2\pi \sigma_i^2} \exp\!\left( -\frac{(x - x_i)^2 + (y - y_i)^2}{2\sigma_i^2} \right)\, dx\, dy \\
&= \frac{1}{4} \left[ \operatorname{erf}\!\left( \frac{X_2 - x_i}{\sqrt{2} \sigma_i} \right) - \operatorname{erf}\!\left( \frac{X_1 - x_i}{\sqrt{2} \sigma_i} \right) \right] \times \left[ \operatorname{erf}\!\left( \frac{Y_2 - y_i}{\sqrt{2} \sigma_i} \right) - \operatorname{erf}\!\left( \frac{Y_1 - y_i}{\sqrt{2} \sigma_i} \right) \right].
\end{aligned}
\end{equation*}
where erf$(x)$ denotes the error function defined by
\begin{align*}
\mathrm{erf}(x) = \frac{2}{\sqrt{\pi}} \int_0^x e^{-t^2} dt.
\end{align*}
If region $R$ is composed of multiple rectangles, we sum this expression over all such tiles. 

The parameters $\nu$ and $\kappa$ are constrained to be non-negative, i.e., $\nu \ge 0$ and $\kappa \ge 0$, when fitting. Once $(\nu, \kappa)$ are estimated, the weighting coefficients $w_i$ can be computed via \eqref{wi}. Furthermore, when evaluating $\eta(m_i)$ in \eqref{ltmxy}, the term  $E(w_i)$ as defined in \eqref{emi}, which is computed over all events up to time $t_i$ with $m \ge m_0$,
\begin{equation*}
E(w_i) = \frac{1}{i} \sum_{k = 1}^{i} w_k,
\end{equation*}
is also obtained.

\subsection{Fitting parameters in EEPAS model}
We estimate the parameters appearing in the EEPAS model \eqref{ltmxy} by maximizing the log-likelihood function:
\begin{equation}\label{lnL}
\begin{aligned}
\ln{\mathcal{L}(\lambda)} \equiv \sum_{\substack{t_j \in [t_s, t_e)\\ m_j \ge m_T\\ (x_j, y_j) \in R}} \ln{\lambda(t_j, m_j, x_j, y_j)} 
- \int_{t_s}^{t_e} \int_{m_T}^{m_u} \iint_{R} \lambda(t, m, x, y)\, dA\, dm\, dt.
\end{aligned}
\end{equation}
While the likelihood targets events with $m \ge m_T$, events with $m \in [m_0, m_T)$ still contribute as precursors in the summation term $\sum_i$ via \eqref{ltmxy}.

The first term is computed exactly as in the PPE section, using the magnitude range $[m_T, m_u]$. For the second term, we first evaluate the time integral $\int_{t_s}^{t_e} f_i(t)\, dt$. The second term in \eqref{lnL}, representing the expected number of events, is expanded as:
\begin{equation*}
\begin{aligned}
\int_{t_s}^{t_e} \int_{m_T}^{m_u} \iint_{R} & \lambda(t, m, x, y)\, dA\, dm\, dt \\
={}& \mu \int_{t_s}^{t_e} \int_{m_T}^{m_u} \iint_{R} \lambda_0(t, m, x, y)\, dA\, dm\, dt \\
& + \sum_{i: t_i < t_e - \text{delay}} \left( \int_{t_s}^{t_e} f_i(t)\, dt \right) \\
& \times \left( \int_{m_T}^{m_u} \frac{\eta(m_i)\, w_i}{\Delta(m)}\, g_i(m)\, dm \right)
\left( \iint_{R} h_i(x, y)\, dA \right).
\end{aligned}
\end{equation*}
Note that the summation runs over all past events $i$ that can contribute to the window $[t_s, t_e]$.

To evaluate the time integral, we consider two cases depending on the time of the $i$-th event. Due to the ${}^{t - \text{delay}}$ convention, the lower limit becomes $\max(t_s,\, t_i + \text{delay})$. Let $s = \log_{10}(t - t_i)$ so that $ds = \frac{dt}{(t - t_i)\ln 10}$, and define $\mu_i = a_T + b_T m_i$.

When $t_i \le t_s - \text{delay}$, the time integral becomes:
\begin{eqnarray}\label{intfi_case1}
\int_{t_s}^{t_e} f_i(t)\, dt 
&=& \int_{\log_{10}(t_s - t_i)}^{\log_{10}(t_e - t_i)} \frac{1}{\sigma_T \sqrt{2\pi}} 
\exp\!\left[-\frac{1}{2} \left( \frac{s - \mu_i}{\sigma_T} \right)^2 \right] ds \notag \\
&=& \frac{1}{2} \left[ \operatorname{erf}\!\left( \frac{\log_{10}(t_e - t_i) - \mu_i}{\sqrt{2}\sigma_T} \right)
- \operatorname{erf}\!\left( \frac{\log_{10}(t_s - t_i) - \mu_i}{\sqrt{2}\sigma_T} \right) \right].
\end{eqnarray}

When $t_s - \text{delay} < t_i < t_e - \text{delay}$, we obtain:
\begin{eqnarray}\label{intfi_case2}
\int_{t_s}^{t_e} f_i(t)\, dt 
&=& \int_{t_i + \text{delay}}^{t_e} \frac{1}{(t - t_i)\ln 10\, \sigma_T \sqrt{2\pi}} 
\exp\!\left[-\frac{1}{2} \left( \frac{\log_{10}(t - t_i) - \mu_i}{\sigma_T} \right)^2 \right] dt \notag \\
&=& \frac{1}{2} \left[ \operatorname{erf}\!\left( \frac{\log_{10}(t_e - t_i) - \mu_i}{\sqrt{2}\sigma_T} \right)
- \operatorname{erf}\!\left( \frac{\log_{10}(\text{delay}) - \mu_i}{\sqrt{2}\sigma_T} \right) \right].
\end{eqnarray}

These two cases can be written in a unified form:
\[
\int_{t_s}^{t_e} f_i(t)\, dt
= \frac{1}{2} \left[
\operatorname{erf}\!\left( \frac{\log_{10}(t_e - t_i) - \mu_i}{\sqrt{2} \sigma_T} \right)
- \operatorname{erf}\!\left( \frac{\log_{10}(\max\{t_s,\, t_i + \text{delay}\} - t_i) - \mu_i}{\sqrt{2} \sigma_T} \right)
\right],
\]
which reduces to \eqref{intfi_case1} or \eqref{intfi_case2} as appropriate. If $t_i \ge t_e - \text{delay}$, then $\int_{t_s}^{t_e} f_i(t)\, dt = 0$.

The magnitude integral
\[
\int_{m_T}^{m_u} \frac{\eta(m_i)}{\Delta(m)}\, g_i(m)\, dm
\]
is computed numerically. Details are provided in Section \ref{sec:optimization}.

For the spatial integral $\iint_R h_i(x, y)\, dA$, direct quadrature over $R$ may be computationally expensive due to the large number of events with $m \ge m_0$. To accelerate computation, we divide $R$ into rectangular subregions $[X_\ell, X_r] \times [Y_d, Y_u]$ and exploit the separability of $h_i(x, y)$:
\begin{eqnarray}
\iint_{R} h_i(x, y)\, dA 
&=& \sum_{\text{subregions}} \left\{ \int_{X_\ell}^{X_r} \frac{1}{\sqrt{2\pi} \sigma_A 10^{\frac{b_A m_i}{2}}} 
\exp\!\left[ -\frac{(x - x_i)^2}{2 \sigma_A^2 10^{b_A m_i}} \right] dx \right. \notag \\
&& \hspace{2cm} \times \left. \int_{Y_d}^{Y_u} \frac{1}{\sqrt{2\pi} \sigma_A 10^{\frac{b_A m_i}{2}}} 
\exp\!\left[ -\frac{(y - y_i)^2}{2 \sigma_A^2 10^{b_A m_i}} \right] dy \right\} \notag \\
&=& \frac{1}{4} \sum_{\text{subregions}} \left[ \operatorname{erf}\!\left( \frac{X_r - x_i}{\sqrt{2}\, \sigma_A 10^{b_A m_i / 2}} \right) 
- \operatorname{erf}\!\left( \frac{X_\ell - x_i}{\sqrt{2}\, \sigma_A 10^{b_A m_i / 2}} \right) \right] \notag \\
&& \times \left[ \operatorname{erf}\!\left( \frac{Y_u - y_i}{\sqrt{2}\, \sigma_A 10^{b_A m_i / 2}} \right) 
- \operatorname{erf}\!\left( \frac{Y_d - y_i}{\sqrt{2}\, \sigma_A 10^{b_A m_i / 2}} \right) \right].
\label{inth1i}
\end{eqnarray}

\subsection{Optimization of EEPAS}

We are now in a position to fit the parameters of the EEPAS model. In our simulations, we consider earthquakes with magnitudes no smaller than $m_0$ and focal depths shallower than 40 kilometers. Throughout the optimization process, all geographic coordinates (longitude and latitude) are projected into kilometers to maintain consistency with the model's spatial kernel.

The initial estimate for the $\Psi$ phenomenon is obtained following the method described in \cite{psi}. For parameter optimization, we apply constrained Nelder-Mead optimization to search for a local minimum of $-\ln \mathcal{L}(\lambda)$ under box constraints. Following the procedure outlined in \cite{2023ER}, we fit the parameters in three steps:

\begin{itemize}
    \item Step 1: Fix the values of $b_M$, $\sigma_M$, $b_T$, $\sigma_T$, and $b_A$, and fit the four parameters $a_M$, $a_T$, $\sigma_A$, and $\mu$.
    \item Step 2: Using the estimates obtained in Step 1, fit $\sigma_M$, $b_T$, $\sigma_T$, $b_A$, and $\mu$, initializing from their Step 1 values while keeping the other four parameters fixed.
    \item Step 3: Jointly fit all eight parameters, namely $a_M$, $a_T$, $\sigma_A$, $\sigma_M$, $b_T$, $\sigma_T$, $b_A$, and $\mu$, while keeping $b_M$ fixed. Initial values for $a_M$, $a_T$, and $\sigma_A$ are taken from Step 1, and the remaining parameters are initialized using the results from Step 2.
\end{itemize}

{\it This three stage fitting procedure is fully configurable in our implementation.} Users may define the number of optimization stages, specify which parameters are fixed or optimized at each stage, and assign bounds and initial values for each parameter. This flexibility enables users to adopt customized parameter grouping strategies, such as those used in \cite{RE,rastin2021space}. Our empirical results suggest that different grouping orders can lead to higher values of the log-likelihood function. Finally, in addition to the default constrained Nelder Mead method, our framework also supports alternative optimization approaches. For further details, see Section \ref{sec:optimization}.

\subsection{Forecasting via EEPAS model}
We are now in a position to discuss the forecast via the optimized EEPAS model. To compute the forecast rate density $\lambda(t, m, x, y)$ in \eqref{ltmxy}, we evaluate integrals over specified time periods $[t_e, t_f)$, magnitude intervals $[M_1, M_2]$, and region $\tilde{S}$. That is, we sum up the total rate density in the domain $[t_e, t_f) \times [M_1, M_2] \times \tilde{S}$:
\begin{align*}
    &\int_{t_e}^{t_f} \! \int_{M_1}^{M_2} \! \iint_S \lambda(t, m, x, y)\, dA\, dm\, dt \\[1ex]
    ={}& \mu \left( \int_{t_e}^{t_f} f_0(t)\, dt \right) \left( \int_{M_1}^{M_2} g_0(m)\, dm \right) 
       \left( \iint_{\tilde{S}} h_0(x, y)\, dA \right) \\[1ex]
    & + \sum_{\substack{t_i \le t_e - \text{delay} \\ m_i \ge m_0}} \eta(m_i)\, w_i 
       \left( \int_{t_e}^{t_f} f_i(t)\, dt \right)
       \left( \int_{M_1}^{M_2} \frac{g_i(m)}{\Delta(m)}\, dm \right)
       \left( \iint_{\tilde{S}} h_i(x, y)\, dA \right).
\end{align*}
Here $h_0(x, y)$ sums the contributions from those earthquakes which occurred in the data collection region $S$ with $t_i < t_e - \text{delay}$ and $m_i \ge m_T$. For convenience, we set $\tilde{S}$ to be a rectangular region $[X_\ell, X_r] \times [Y_d, Y_u]$. The integrals of $f_0, g_0, f_i, h_i$ are computed explicitly as
\begin{align*}
&\int_{t_e}^{t_f} f_0(t)\, dt = \ln(t_f - t_0) - \ln(t_e - t_0), \\
&\int_{M_1}^{M_2} g_0(m)\, dm = \exp\!\left[-\beta(M_1 - m_T)\right] - \exp\!\left[-\beta(M_2 - m_T)\right], \\
&\int_{t_e}^{t_f} f_i(t)\, dt 
 = \frac{1}{2} \left[ 
  \operatorname{erf}\!\left( \frac{\log_{10}(t_f - t_i) - a_T - b_T m_i}{\sqrt{2}\, \sigma_T} \right)
- \operatorname{erf}\!\left( \frac{\log_{10}(t_e - t_i) - a_T - b_T m_i}{\sqrt{2}\, \sigma_T} \right) 
\right],
\end{align*}
and
\begin{align*}
\iint_{\tilde{S}} h_i(x, y)\, dA 
&= \frac{1}{4} 
\left[ 
  \operatorname{erf}\!\left( \frac{X_r - x_i}{\sqrt{2}\, \sigma_A 10^{\frac{b_A m_i}{2}}} \right) 
- \operatorname{erf}\!\left( \frac{X_\ell - x_i}{\sqrt{2}\, \sigma_A 10^{\frac{b_A m_i}{2}}} \right) 
\right] \\
&\quad \times 
\left[ 
  \operatorname{erf}\!\left( \frac{Y_u - y_i}{\sqrt{2}\, \sigma_A 10^{\frac{b_A m_i}{2}}} \right) 
- \operatorname{erf}\!\left( \frac{Y_d - y_i}{\sqrt{2}\, \sigma_A 10^{\frac{b_A m_i}{2}}} \right) 
\right].
\end{align*}
The integrals $\iint_{\tilde{S}} h_0(x, y)\, dA$ and $\int_{M_1}^{M_2} \frac{g_i(m)}{\Delta(m)}\, dm$ are computed numerically.

\section{The Framework}

\begin{figure}
    \centering
    \includegraphics[width=1\linewidth]{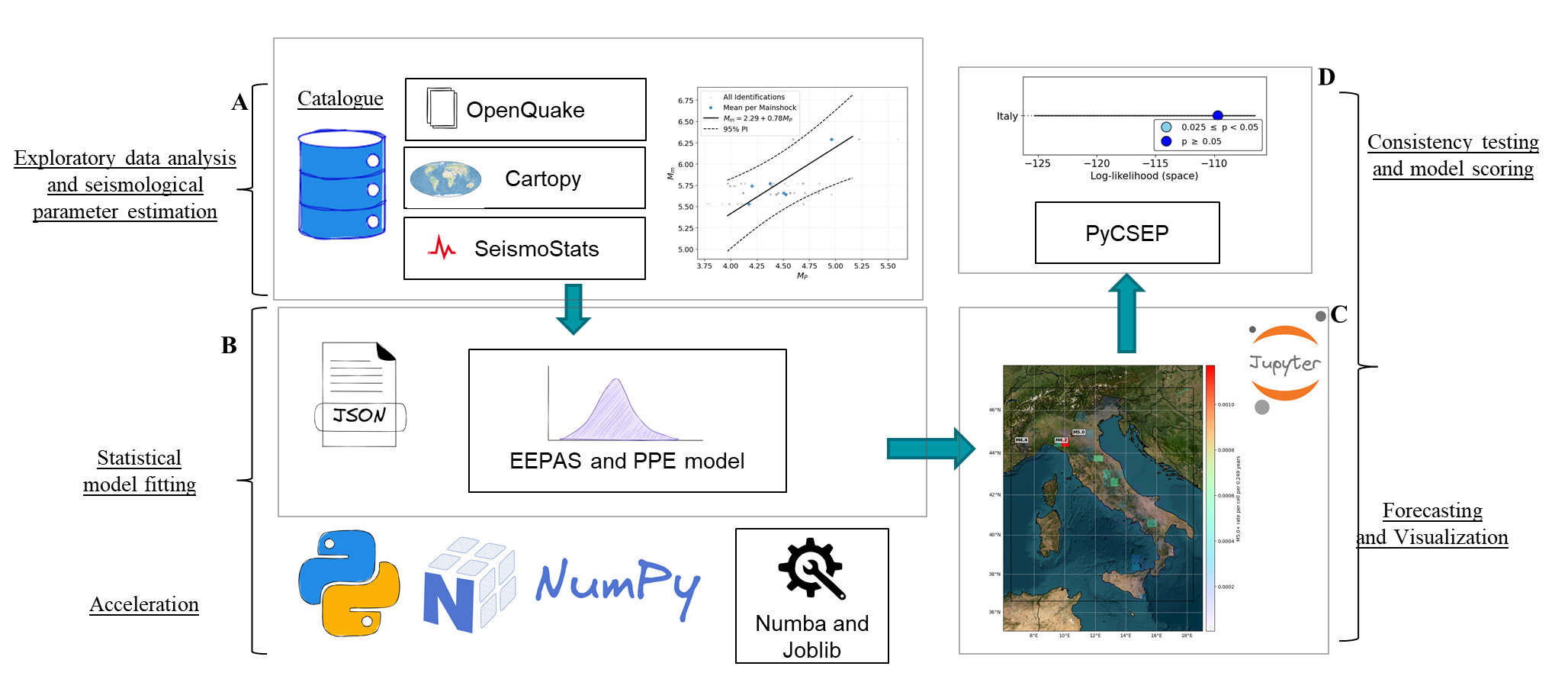}
    \caption{The proposed framework and software stack. (a) Perform exploratory data analysis and estimate seismological parameters. The framework integrates seamlessly with parameter estimation packages and is built upon widely used plotting libraries. (b) Fit the EEPAS and PPE models. Model fitting and forecasting are accelerated using vectorization, parallelization, and just-in-time (JIT) compilation. (c) Generate forecasts and conduct analysis and visualization. (d) The module can be seamlessly integrated with popular statistical testing tools.}
    \label{fig:framework}
\end{figure}

The framework is illustrated in \Cref{fig:framework}. It is built upon Python and the \texttt{SciPy} ecosystem \cite{virtanen2020scipy}, allowing seamless integration with visualization tools such as \texttt{Matplotlib} \cite{Hunter:2007} and \texttt{CartoPy} \cite{Cartopy}, and enabling interactive analysis within \texttt{Jupyter Notebook} \cite{kluyver2016jupyter}. In addition, we leverage numerical packages such as \texttt{NumPy} \cite{harris2020array}, \texttt{Numba}, and \texttt{Joblib} to support vectorization, parallelization, and just-in-time (JIT) compilation. 

All model parameters, file paths, and runtime settings are controlled through a centralized JSON configuration file. This eliminates hard-coded values and allows users to easily adapt the model to different datasets, initialization strategies, or optimization algorithms without modifying the source code. The code is organized into distinct, well-defined modules, enhancing readability, maintainability, and extensibility for both developers and end users.

The framework is designed to integrate seamlessly with other essential components of the earthquake forecasting workflow. It provides connectors to preprocessing tools such as \texttt{SeismoStats} \cite{mirwald2025seismostats} for estimating the $b$-value in the Gutenberg–Richter law, and to post-analysis libraries such as \texttt{pyCSEP} for statistical testing and model comparison. This enables a complete, end-to-end research pipeline—from catalog preparation and model construction to evaluation and benchmarking. The entire codebase is open-source and extensively documented, with detailed explanations of each module's functionality, inputs, and configurable parameters.

\begin{figure}
\centering
\includegraphics[width=.9\textwidth]{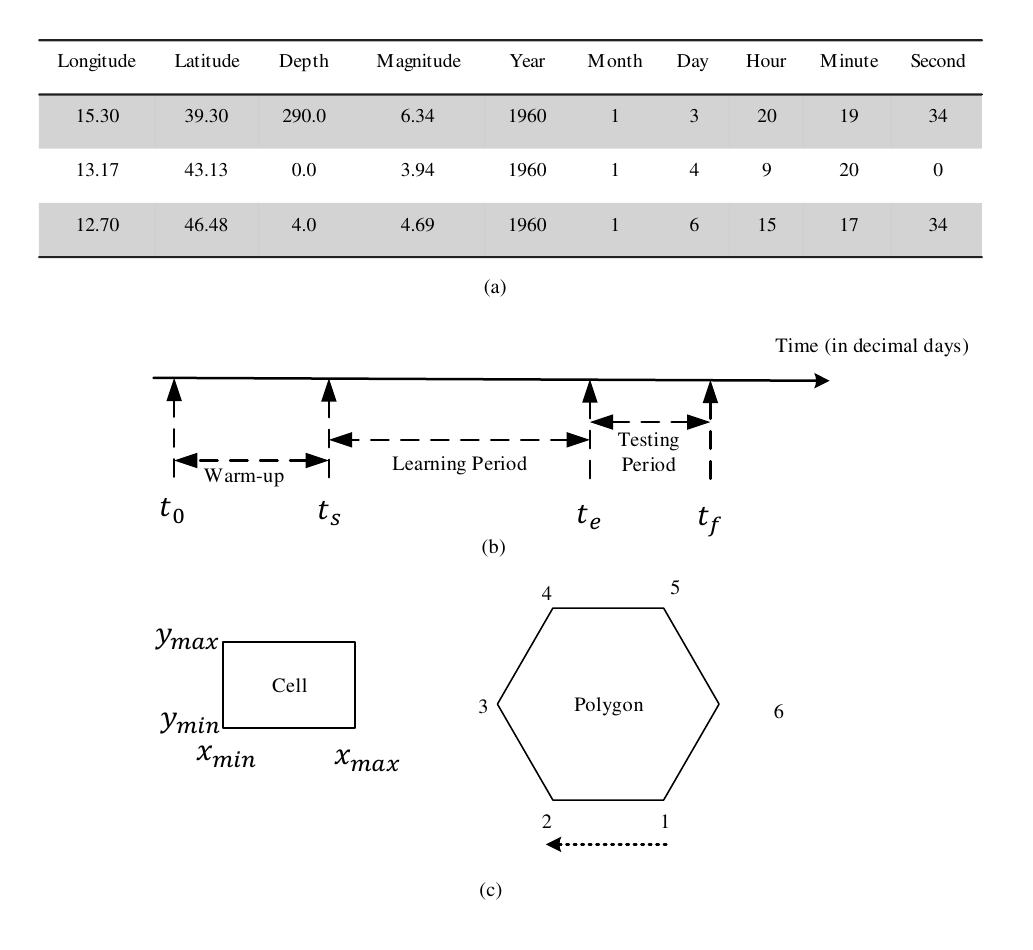}
\caption{(a) Example of catalog data. (b) Diagram illustrating warm-up, learning, and testing periods. (c) Definition of the cell-based or polygonal two-dimensional study region.}
\label{fig:format}
\end{figure}

\subsection{Data Preparation and Seismological Parameter Estimation}

To initiate the workflow, the user inputs a standard earthquake catalog containing the essential event data. Let the $i$-th earthquake (indexed $i = 1, \dots, N_{\text{total}}$), observed in a data collection region $S$, occur at time $t_i$ with magnitude $m_i \geq m_0$ and location $(x_i, y_i)$. A typical catalog structure is shown in \Cref{fig:format}(a). Our framework supports widely used formats, including \texttt{ZMAP} \cite{wiemer2001software}, \texttt{pyCSEP}, and \texttt{SeismoStats}.

The catalog is divided into three consecutive intervals: a warm-up period, a learning period, and a testing period, as illustrated in \Cref{fig:format}(b). The user specifies the start and end times of the learning period, denoted by $t_s$ and $t_e$, respectively. Regarding the temporal boundaries, $t_0$ is defined as the start time of the data collection (ensuring $t_0 < t_1$ to avoid singularities in temporal kernels), and $t_f$ is set to the end time of the catalog. The warm-up period $[t_0, t_s)$ provides critical history for model stability. Specifically, earthquakes with $m > m_0$ during this phase serve as precursors for the time-varying component of EEPAS, while those with $m > m_T$ inform the background component of the PPE model.

To specify the study region, the geographical setup includes two regions: the neighborhood region $S$ and the testing region $R$, both defined using either a rectangular grid or an irregular polygonal boundary, as shown in \Cref{fig:format}(c). For cell-based regions, the user must specify the bounding box using $x_{\min}, x_{\max}, y_{\min}, y_{\max}$. For polygonal regions, the area is defined by the coordinates of the polygon’s vertices in clockwise order, with no repetitions. The neighborhood region $S$ must strictly contain the testing region $R$ to avoid boundary effects, such as the truncation of precursor events outside $R$ that may influence target events near the edge. Similarly, the difference between $m_T$ and $m_0$ should be at least 2.0. This follows from the first $\Psi$ scaling relation, which suggests that the largest precursors are typically one magnitude unit smaller than the mainshock, with many being even smaller \cite{rhoades202220}. To support this setup, we provide a robust projection module that performs geographically appropriate transformations. This replaces naive latitude and longitude to kilometer conversions, ensuring that coordinates are suitable for subsequent spatial analysis.

Once the data are prepared and the relevant parameters and regions are specified, users can apply \texttt{SeismoStats}, which offers various $b$-value estimators for quantifying the relative frequency of small versus large earthquakes. Additionally, we provide a fully automated pipeline that takes raw earthquake catalogs as input and outputs initial parameter estimates for initializing the EEPAS model. Specifically, the rectangular algorithm described in \cite{psi} has been implemented to detect the $\Psi$ phenomenon, identifying all sets of precursory variables $\{M_p, T_p, A_p\}$ for each mainshock. Deriving the $\Psi$ scaling relations requires addressing a significant statistical challenge inherent in the data structure. The dataset contains multiple $\Psi$ identifications (observations) nested within each mainshock (group), introducing a within-group negative correlation between precursory area ($A_P$) and precursory time ($T_P$), known as the space-time trade-off. A naive Ordinary Least Squares (OLS) regression applied to the pooled data fails to distinguish this \textit{within-mainshock} trade-off from the true \textit{between-mainshock} scaling relationship of interest, resulting in biased and inconsistent estimators.

To resolve this issue, a two-stage estimation procedure is employed. First, the common $A_P$–$T_P$ trade-off slope ($b$) is estimated using a fixed-effects (or “within”) estimator. This model isolates the variation \textit{within} each mainshock by demeaning the variables relative to their group-specific means, providing a robust estimate of the shared trade-off slope across mainshocks. In the second stage, this estimated slope is used to remove the confounding effect. Rather than relying on simple arithmetic means, which would be biased, we compute projected representative points for each mainshock. This statistical projection adjusts each mainshock’s mean $\log A_P$ and $\log T_P$ along the common trade-off slope ($b$) to a single reference point, such as the global mean $\overline{\log A_P}$ or $\overline{\log T_P}$. The result is a purified dataset in which each mainshock is represented by a single observation corrected for the internal trade-off. In the third step, standard OLS regression is applied to this projected dataset to obtain unbiased estimates of the true \textit{between-mainshock} scaling relationships among $\log A_P$, $\log T_P$, and $M_P$. The resulting regression coefficients serve as the initial parameter estimates. For example, in estimating the time–magnitude relation $\log T_P = a_T + b_T M_P$, the initial values for the intercept ($a_T$) and slope ($b_T$) are taken directly from the regression output, while the standard deviation $\sigma_T$ is set to the residual standard error, which quantifies the spread of residuals around the fitted line.

Finally, users can perform exploratory data analysis and visualization using modules built on top of \texttt{Matplotlib} and \texttt{CartoPy}, and conduct interactive analysis within \texttt{Jupyter Notebook}.

\subsection{Model Fitting and Acceleration} \label{sec:optimization}

Acceleration is achieved through a multi-faceted approach targeting key computational bottlenecks, primarily by leveraging analytical simplifications, vectorization, and parallelization.

Core numerical loops are first optimized using \texttt{NumPy} vectorization and \texttt{Numba}’s Just-In-Time (JIT) compilation. JIT compilation significantly speeds up computationally intensive low-level loops when operations are implemented with \texttt{NumPy}. For example, the log-likelihood evaluation involves deeply nested loops over thousands of events, which is inefficient in pure Python. By implementing these calculations using \texttt{NumPy} and applying \texttt{Numba}’s \texttt{@njit} decorator, the functions are compiled into efficient machine code, substantially reducing runtime, as illustrated in \Cref{lst:numba}.

\begin{figure}
\begin{lstlisting}[style=mypython]
from numba import njit, prange
import numpy as np

@njit(parallel=True, fastmath=True)
def compute_event_terms_fast(num_targets):
    log_lambda_values = np.zeros(num_targets)
    for i in prange(num_targets):
        # Complex calculations for each event
        log_lambda_values[i] = np.log(i + 1.0)
    return np.sum(log_lambda_values)
\end{lstlisting}
\caption{Accelerating event term computation using Numba}
\label{lst:numba}
\end{figure}

Process-based parallelism is also used to distribute independent high-level tasks across multiple CPU cores. This is especially effective when the implementation depends on pure Python or SciPy routines. For example, the forecast module requires performing hundreds of independent 2D numerical integrals (one per grid cell), each of which is computationally expensive. A serial implementation is prohibitively slow. We employ Joblib to parallelize such embarrassingly parallel tasks. With \texttt{joblib.Parallel}, independent calls to \texttt{dblquad} are distributed across CPU cores, bypassing Python’s Global Interpreter Lock (GIL), as shown in \Cref{lst:joblib}.

\begin{figure}
\begin{lstlisting}[style=mypython]
from joblib import Parallel, delayed
from scipy.integrate import dblquad

def compute_cell_integral(i):
    result, _ = dblquad(...)
    return result

ExpE = np.array(Parallel(n_jobs=-1)(
    delayed(compute_cell_integral)(i)
    for i in range(celln)
))
\end{lstlisting}
\caption{Parallelizing integrals using \texttt{joblib}}
\label{lst:joblib}
\end{figure}

At a lower level, algorithmic substitution was employed to replace numerical integration with analytical solutions wherever possible. For example, one- and two-dimensional Gaussian integrals are evaluated using the error function, as discussed previously. Although \texttt{erf} is itself defined via an integral, standard libraries implement it using fast polynomial approximations or rational expansions. As a result, replacing numerical quadrature with \texttt{erf} calls offers substantial performance gains.

For integrals that cannot be evaluated analytically — such as the spatial kernel integral $\iint_R h_{0i}(x,y)\,dA$ and the EEPAS magnitude integral
\[
\int_{m_T}^{m_u} \frac{\eta(m_i)}{\Delta(m)}\, g_i(m)\,dm,
\]
a hybrid integration strategy is adopted. Two modes are available. In \emph{accuracy mode}, adaptive quadrature is applied using SciPy’s \texttt{dblquad} or the vectorized \texttt{quad\_vec} routine, when vectorization is feasible. The \texttt{quad\_vec} algorithm adaptively subdivides subintervals with the largest estimated errors, enabling partial parallelization while preserving accuracy. In \emph{fast mode}, fixed-grid trapezoidal integration with user-defined resolution is used. This method is considerably faster and can be further accelerated with Numba.

The trade-off between accuracy and speed is user-configurable. For instance, during the learning phase, fixed-grid integration is typically sufficient for smooth normalization terms and intermediate optimization steps where ultra-high precision is not required. In contrast, during the forecasting phase, adaptive methods may be preferable for final evaluations where numerical precision is critical, albeit with greater computational cost.



On the other hand, in constrained optimization, if the true optimal value of a parameter lies outside the predefined search space, the optimizer can only converge to a suboptimal point on the boundary, resulting in biased parameter estimates. To address this issue, we implement an automated workflow that iteratively adjusts parameter boundaries. The procedure is as follows: run the full EEPAS optimization; identify parameters that have converged too close to their specified bounds; if any such cases are detected, automatically expand the corresponding boundaries and re-run the optimization.  A parameter is considered to be hitting a boundary if its estimated value lies within a small fraction (default: 1\%) of the total boundary range. For small-valued parameters, such as variances, a more robust absolute tolerance is used instead. When a problematic boundary is detected, it is widened by an expansion factor (default: 2.0), while strictly respecting the physical constraints of the EEPAS model. In particular, variance terms and slope parameters are constrained to remain strictly positive (e.g., lower bounds floored at $10^{-6}$), and the proportion parameter $u$ is constrained to the interval $[0,1]$. The boundary adjustment loop terminates when no parameters are hitting their bounds, or when the improvement in the negative log-likelihood (NLL) becomes negligible.

By default, we use the constrained Nelder–Mead simplex algorithm as the primary optimizer. However, users are free to select alternative methods, such as the Truncated Newton Constrained (TNC) algorithm, the Limited-memory Broyden–Fletcher–Goldfarb–Shanno algorithm with bound constraints (L-BFGS-B), or the Sequential Least Squares Quadratic Programming (SLSQP) algorithm. In addition, users may enable Basin-Hopping or multistart optimization strategies to escape local minima.

\subsection{Forecasting}
Our platform implements a Pseudo-prospective Mode to generate forecasts, simulating a realistic operational environment. This approach rigorously evaluates the model by applying it to historical data that were not used during calibration \cite{rhoades2011, rhoades202220, 2023ER}. A core principle of this mode is its one-time calibration strategy: model parameters are calibrated once using a designated historical learning period and are then held fixed throughout the entire test period. As emphasized by \cite{rhoades202220}, this fixed-parameter approach ensures forecast stability and prevents overfitting to the test data, enabling a robust assessment of predictive performance.

Forecast generation operates on a three-month rolling mechanism, which balances parameter stability with the incorporation of new seismic data. While model parameters remain fixed, the forecasts are updated by including newly observed earthquakes from the test period as they occur, after a processing delay. Specifically, an initial three-month forecast is generated using the calibrated parameters. For each subsequent three-month window, the expected rates ($\lambda$) are recomputed by integrating the newly occurred events. This rolling process allows the model to account for the dynamic influence of recent seismicity, which is crucial for capturing precursory effects \cite{rhoades2011}, while maintaining the forecast stability advocated by \cite{christophersen2010}.

Finally, the platform produces forecasts in the form of a gridded base forecast. Predictions are discretized into spatial cells of $0.1^\circ \times 0.1^\circ$ (longitude and latitude) and magnitude bins of width 0.1. The final output specifies the expected number of earthquakes within each space–magnitude bin, providing a standardized format suitable for evaluation.

\subsection{Tests for Grid-Based Forecasts}

Our platform streamlines forecast evaluation by generating gridded base forecasts that are directly compatible with the \texttt{pyCSEP} toolkit. This seamless integration allows users to rigorously assess forecast performance using \texttt{pyCSEP}'s comprehensive suite of statistical tests.

These evaluations are broadly categorized into consistency tests and comparative scoring rules. Consistency tests evaluate whether the observed seismicity is statistically consistent with the forecast. This includes fundamental checks such as the log-likelihood test (L-test) and the number test (N-test), both of which can be applied using either Poisson or Negative Binomial distributions. The latter accommodates the overdispersion commonly observed due to earthquake clustering. \texttt{pyCSEP} also supports domain-specific tests, including the spatial test (S-test) and the magnitude test (M-test), which assess whether the forecast accurately captures the spatial and magnitude distributions of observed events.

Beyond consistency, \texttt{pyCSEP} facilitates the comparison and ranking of different models through scoring rules. These include the log-likelihood score for overall model fit, the Brier score \cite{glenn1950verification}, which is particularly suitable for evaluating rare events, and the Kagan information score \cite{kagan2009testing}, which quantifies the forecast's spatial informativeness relative to a reference model.

By producing \texttt{pyCSEP}-ready outputs, our platform enables researchers to efficiently validate, compare, and refine their earthquake forecast models using established statistical benchmarks.

\begin{figure}
    \centering
    \includegraphics[width=0.5\linewidth]{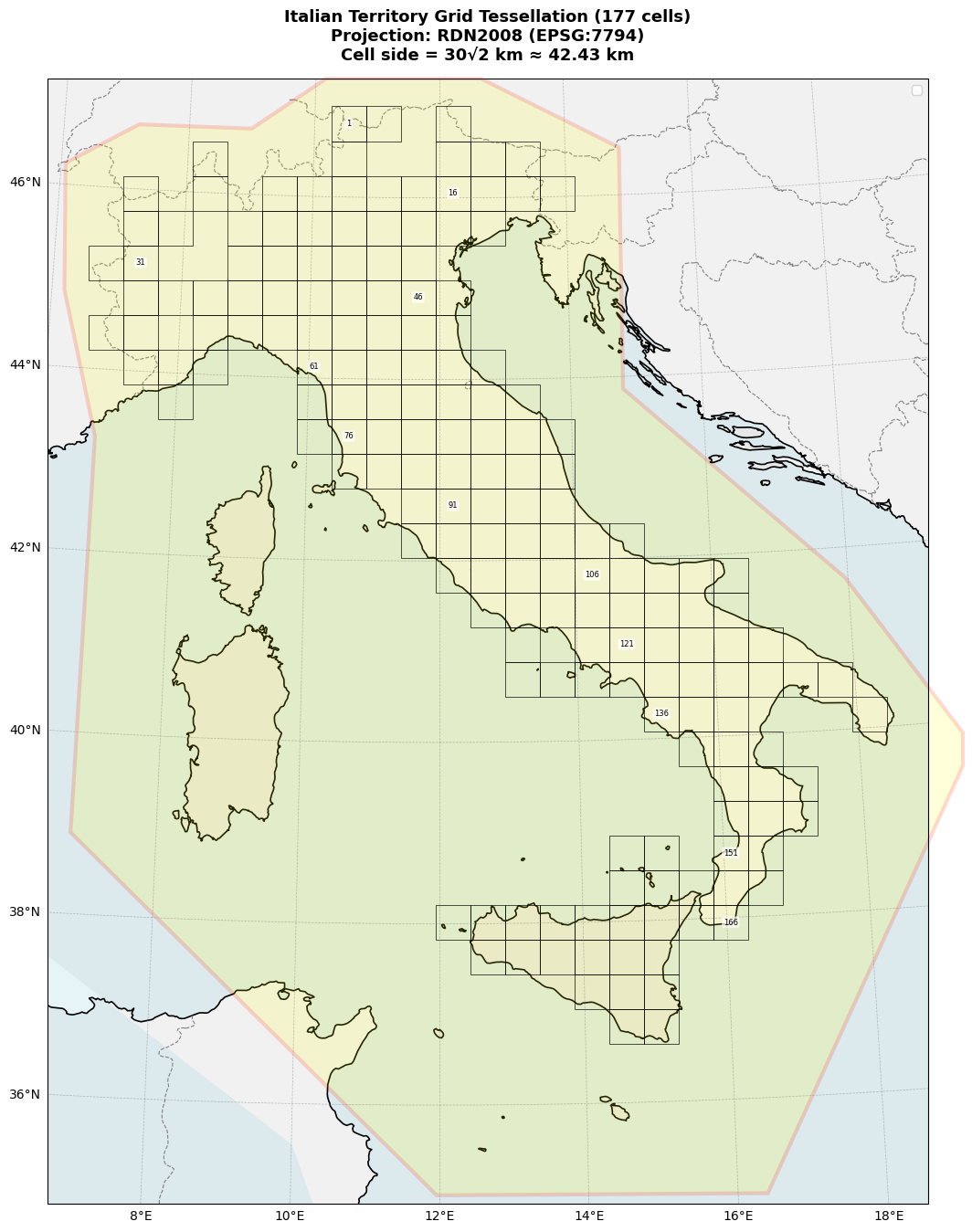}
    \caption{The testing region ($R$) within the larger neighborhood region ($S$).}
    \label{fig:italy_region}
\end{figure}

\section{Evaluations}

\subsection{Reproducing the Results from Literature}

To demonstrate the efficacy of our framework in real-world applications and its ability to reproduce results from the literature, we first apply it to the Italy region and attempt to replicate the results presented in \cite{2023ER}. Following \cite{2023ER}, the testing region $R$ consists of a regular grid of $30\sqrt{2}$ km square cells covering Italy. Only those cells that contain at least one earthquake with $M \ge 4.0$ occurring inland between 1600 and 1959 (according to the CPTI15 catalog \cite{rovida2020italian}) or between 1960 and 2021 (according to the Homogenized Instrumental Seismic [HORUS] catalog \cite{lolli2020homogenized}) are included, in order to ensure catalog completeness. Cells that are not contiguous with the main analysis polygon are also excluded, yielding a total of 177 square cells. The neighborhood region $S$ is defined by the CPTI15 polygon, as in \cite{rovida2020italian}, to avoid edge effects. These two regions are illustrated in \Cref{fig:italy_region}.

The HORUS catalog from 1960 to 2021 is used in the following analysis. The learning period spans from 1990 to 2011 for fitting the EEPAS model, while the pseudo-prospective testing period is from 2012 to 2021. Consistent with \cite{2023ER}, the minimum magnitude threshold is set to $m_0 = 2.45$, the target magnitude threshold to $m_T = 4.95$, and the delay time to 50 days. Only shallow earthquakes with depth $Z \le 40$ km are included. 

Other seismological parameters are also set to match those in \cite{2023ER}: the $b$-value in the Gutenberg–Richter law is fixed at 1.084; the Omori–Utsu parameters are set to $p = 1.2$ and $c = 0.03$; the standard deviation $\sigma_U$ used in the aftershock model is set to 0.006; and the Bath law offset $\delta$ is set to 0.7. The initial values for the $\Psi$ phenomenon are also adopted from \cite{2023ER}, as shown in \Cref{tbl:reproducing}. Finally, longitude and latitude coordinates are projected into the RDN2008 Italy Zone (East–North) system using our projection module.

\begin{table}
\centering
\caption{Reproduction of the results from \cite{2023ER} using the same parameter settings and reporting the resulting optimized values. Parameters are set according to \cite{2023ER} whenever explicitly specified; otherwise, default values are taken from the original MATLAB implementation. During optimization, the magnitude–frequency slope parameter $b_m$ is fixed at 1.} \label{tbl:reproducing}. 
\begin{tabular}{lrrrr}
\toprule
\textbf{Parameter} & \textbf{Bounds} & \textbf{Initial values} & \textbf{Our results} & \textbf{\cite{2023ER}} \\
\midrule
$a_m$         & 1.0--2.0 & 1.50 & 1.23                 & 1.23                    \\
$b_m$         & - & 1.00 & 1.00                  & 1.00                    \\
$\sigma_m$    & 0.2--0.65 & 0.32 & 0.24                  & 0.24                    \\
$a_t$         & 1.0--3.0 & 1.50 & 2.59                 & 2.71                    \\
$b_t$         & 0.3--0.65 & 0.40 & 0.35                  & 0.32                    \\
$\sigma_t$    & 0.15--0.6 & 0.23 & 0.15                  & 0.15                    \\
$b_a$         & 0.2--0.6 & 0.35& 0.50                  & 0.51                    \\
$\sigma_a$    & 1.0--30.0 & 2.0 & 1.00                  & 1.00                    \\
$u$           & 0--1 & 0.2 & 0.17                  & 0.16                    \\
$a$           & $\ge 0$& 0.005 & 0.62                  & 0.62               \\
$d$           & $\ge 1$& 10.0 & 29.64                 & 30                   \\
$s$           & $\ge 1e-15$& 0.1 & 1e-15                  & $9\times1e-13$                   \\
$\nu$           & 0--1& 0.5 & 0.58                  & 0.61                    \\
$\kappa$          & $\ge 0$ & 0.1 & 0.21                  & 0.13	                    \\
\bottomrule
\end{tabular}
\end{table}

We first perform PPE learning and apply joint optimization to estimate the parameters $a$, $d$, and $s$, as summarized in \Cref{tbl:reproducing}. The optimized values of $a$ and $d$ closely match those reported in \cite{2023ER}, while $s$ converges to a very small value near its lower bound. Note that the triple integral of $\lambda_0$ over a given time period and specific region is mathematically expected to equal the number of earthquakes with $M \ge 4.95$ that occurred in that region during the same period. In our experiments, the triple integral of $\lambda_0$ over the learning period using the optimized $a$, $d$, and $s$ yields a value of 27, which is consistent with the number of observed target events with $M \ge 4.95$ from 1990 to 2011. The resulting log-likelihood is $-514.105$, which closely aligns with the reported value of $-514.11$ in the original experiments.

Next, we use the fitted values of $a$, $d$, and $s$ to estimate the aftershock model by maximizing the likelihood of earthquakes with $m \ge m_T$ occurring within region $R$ during the same learning period (1990 to 2011). The optimized value of $\nu$, representing the proportion of earthquakes that are not aftershocks, is 0.58, which closely matches the value 0.61 reported in \cite{2023ER}. The normalization constant $\kappa$ shows a slight deviation from the original result, which may be attributed to minor differences in the filtering criteria applied to the HORUS dataset that were not explicitly reported, or to unmentioned initialization settings in the original implementation.

We then use the parameters $a$, $d$, $s$, $\nu$, and $\kappa$ to compute the weights for each earthquake and proceed with EEPAS model fitting. Following the three-stage fitting procedure described in \cite{2023ER}, we obtain the final parameter estimates shown in \Cref{tbl:reproducing}. Again, all parameter values are in close agreement with those from the original experiment. The largest deviations occur in $b_T$ and $a_T$, but these remain within 5 to 10 percent. The resulting log-likelihood is $-495.394$, which also closely matches the original value of $-496.06$.

\begin{figure}
    \centering
    \begin{subfigure}[b]{0.45\textwidth}
        \centering
        \includegraphics[width=\linewidth]{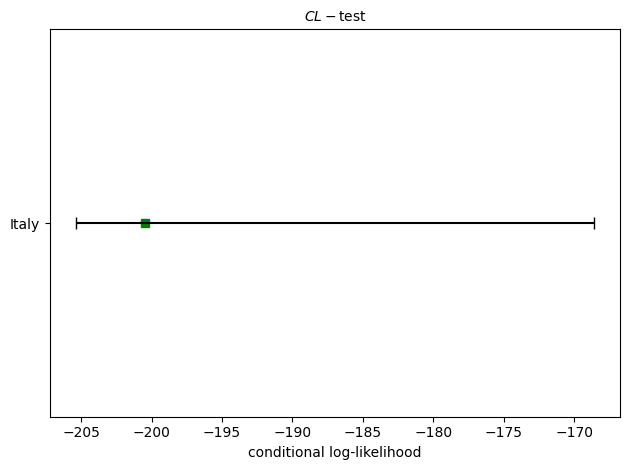}
        \caption{cL-test}
        \label{fig:italy_cltest}
    \end{subfigure}
    \hfill 
    \begin{subfigure}[b]{0.45\textwidth}
        \centering
        \includegraphics[width=\linewidth]{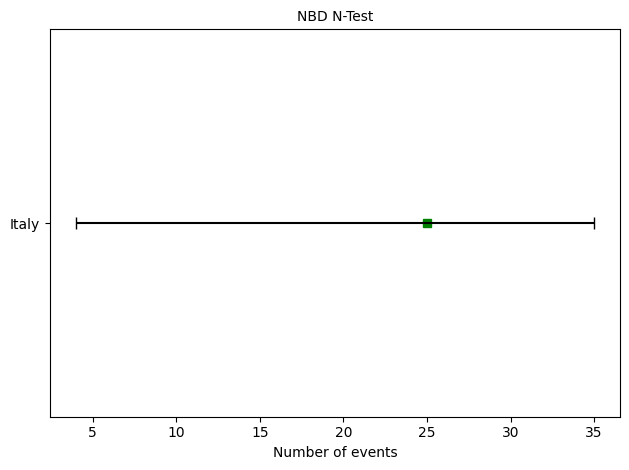}
        \caption{N-test}
        \label{fig:italy_ntest}
    \end{subfigure}
   
    \begin{subfigure}[b]{0.45\textwidth}
        \centering
        \includegraphics[width=\linewidth]{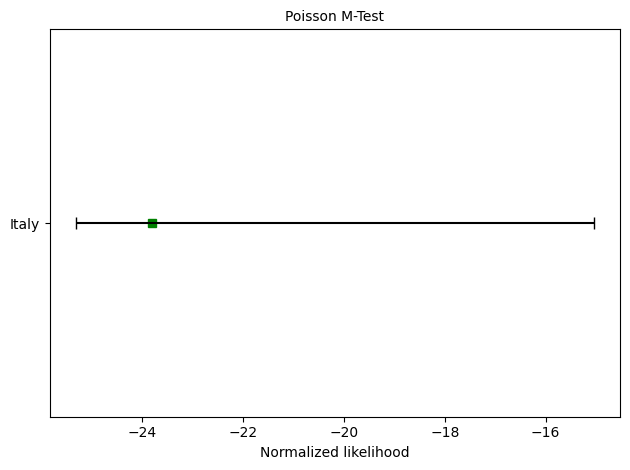}
        \caption{M-test}
        \label{fig:italy_mtest}
    \end{subfigure}
    \hfill 
    \begin{subfigure}[b]{0.45\textwidth}
        \centering
        \includegraphics[width=\linewidth]{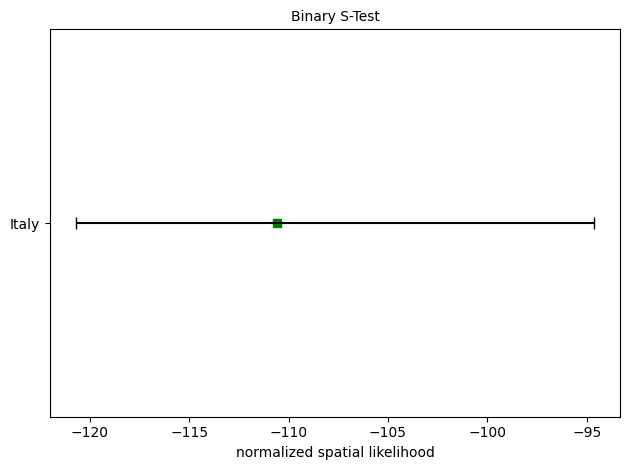}
        \caption{S-test}
        \label{fig:italy_stest}
    \end{subfigure}
    \caption{The results of the consistency tests in the testing set (2012–2021). 
    (a) Conditional likelihood consistency test (cL-test). 
    (b) Number consistency test (N-test). 
    (c) Magnitude consistency test (M-test). 
    (d) Spatial consistency test (S-test).}
    \label{fig:italy_paper_tests}
\end{figure}

We then set the prediction intervals to 3 months and generate the 3-month rolling forecasts in gridded forecast format. A set of new CSEP consistency tests, based on a binary likelihood function, is applied to the forecasts during the testing period. For the $\sigma$ parameter in the Negative Binomial Distribution (NBD), we use $\sqrt{67.76}$, consistent with the setting in \cite{2023ER}. The results are shown in \Cref{fig:italy_paper_tests}. As observed, the output passes all consistency tests, including the Negative Binomial N-test, the binary conditional likelihood (CL) test, and the binary spatial (S) test. Since the model assumes a Gutenberg–Richter frequency–magnitude distribution, it also passes the Poisson M-test.

Overall, this experiment confirms that our framework can be readily used to reproduce results reported in the literature when provided with the same initial parameters. Furthermore, the entire pipeline completes within one hour on a typical notebook equipped with 8 CPU cores, demonstrating the framework’s computational efficiency.

\begin{figure}
    \centering
    
    \begin{subfigure}[b]{0.32\textwidth}
        \centering
        \includegraphics[width=\linewidth]{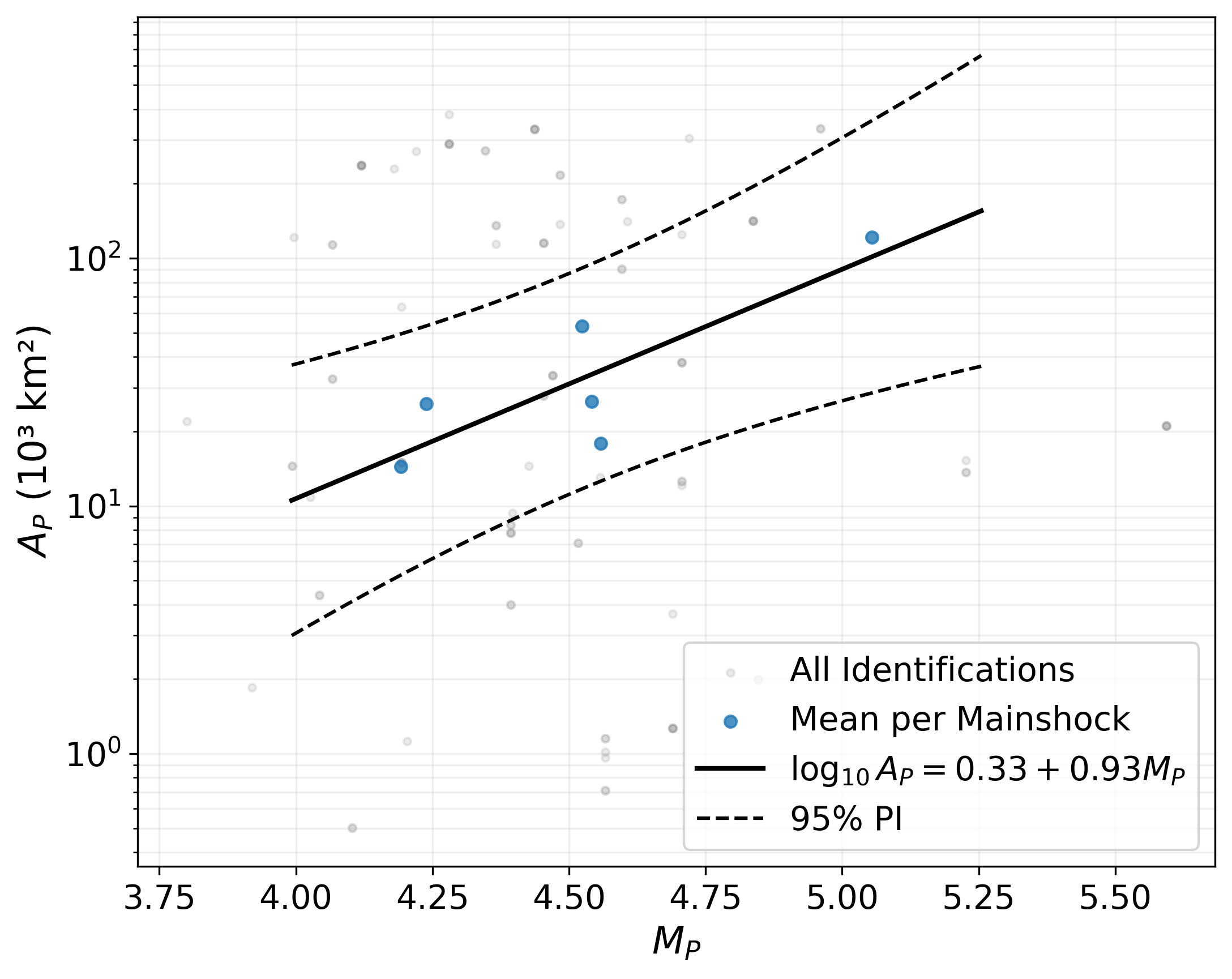}
        \caption{$A_P$ versus $M_p$}
        \label{fig:ApMp}
    \end{subfigure}
    \hfill 
    \begin{subfigure}[b]{0.32\textwidth}
        \centering
        \includegraphics[width=\linewidth]{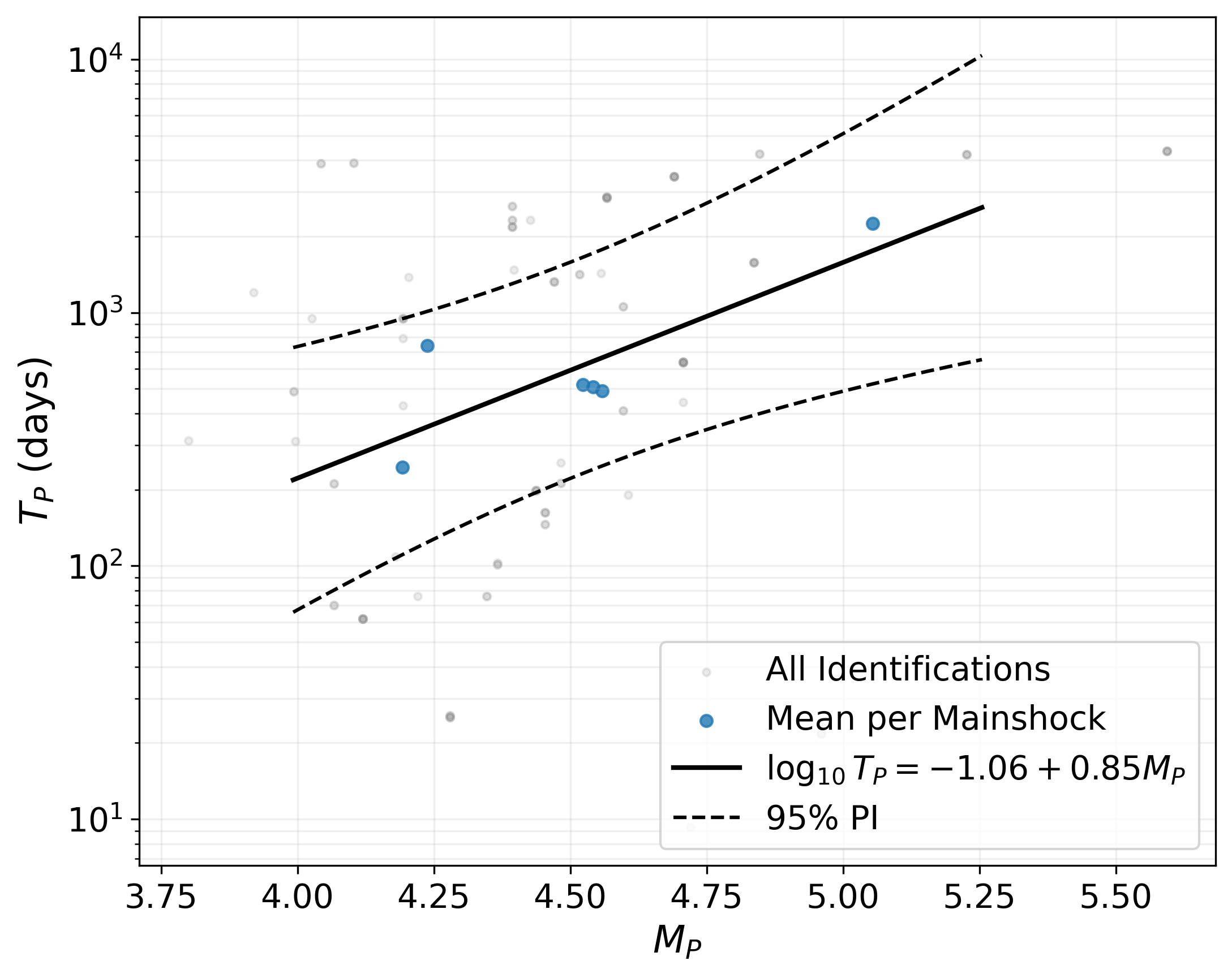}
        \caption{$T_P$ versus $M_p$}
        \label{fig:TpMp}
    \end{subfigure}
    \hfill 
    \begin{subfigure}[b]{0.32\textwidth}
        \centering
        \includegraphics[width=\linewidth]{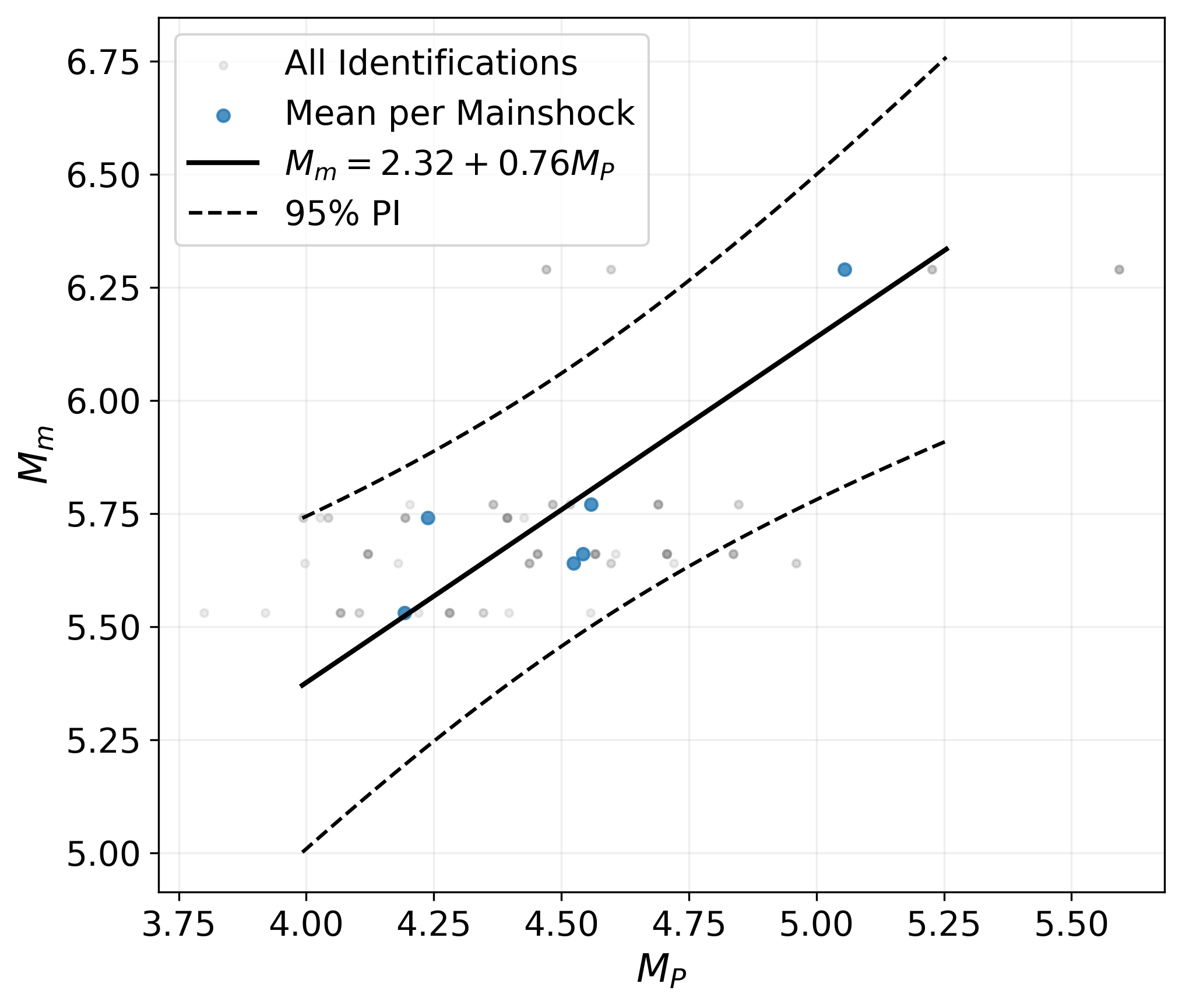}
        \caption{$M_m$ versus $M_p$}
        \label{fig:MmMp}
    \end{subfigure}
    \caption{Scatterplot and fitted scaling relations derived from the algorithmic identification of the $\Psi$ phenomenon using the rectangular algorithm. Each point represents a precursor–mainshock pair, and the fitted lines illustrate the empirical scaling relationships between precursory time, area, and mainshock magnitude.}
    \label{fig:scaling_italy}
\end{figure}

\subsection{An End-to-End Workflow}

Here, we conduct a full end-to-end analysis for the Italy region. The study regions $R$ and $S$, the learning and testing periods, the earthquake catalog, and the depth filtering criteria are identical to those described in the previous subsection. The target magnitude is set to $m_T = 4.95$. Following general recommendations from \cite{rhoades202220}, we set the minimum magnitude threshold to $m_0 = 2.95$, which is approximately two units below $m_T$.

After fixing the analysis parameters, we estimate the $b$-value using the b-more-positive method \cite{lippiello2024b} via the \texttt{SeismoStats} package, yielding a value of 1.013. The values for $\sigma_U$, $p$, $c$, and the delay parameter are kept consistent with those used in the previous experiments. To initialize the EEPAS model, we replace the manual and labor intensive $\Psi$ identification process described in \cite{2023ER} with our automated pipeline, which uses a search radius of 400 km. The resulting $\Psi$ scaling relations are shown in \Cref{fig:scaling_italy}, where the $R^2$ values for area, time, and magnitude are 0.70, 0.68, and 0.77, respectively. These values indicate strong empirical correlations. Initial parameter values are obtained from the slopes, intercepts, and residual standard errors of the linear regressions, as summarized in \Cref{tbl:parameters_opt}.

We then perform PPE learning to estimate the parameters $a$, $d$, and $s$, and verify that the triple integral of $\lambda_0$ over the learning period matches the observed number of events. Aftershock fitting is subsequently performed to estimate the parameters $\nu$ and $k$. Finally, we carry out EEPAS fitting using the previously estimated parameters and initial values. These results are also reported in \Cref{tbl:parameters_opt}. We also enable automatic boundary adjustment during optimization, which ensures that the optimal solution typically lies within the interior of the feasible parameter space rather than on its boundary. The resulting log-likelihood is $-484.23$, which improves upon the $-495.394$ obtained in the previous subsection. Note that since we are maximizing the log-likelihood, a less negative value indicates a better model fit, thereby confirming the improved efficacy of the full pipeline.

\begin{table}
\centering
 \caption{Initial and optimized parameter values for the end-to-end pipeline. Initial EEPAS parameters are computed using the rectangular algorithm and estimated via fixed-effects regression.}  \label{tbl:parameters_opt}
\begin{tabular}{lrr}
\toprule
\textbf{Parameter} & \textbf{Initial Value} & \textbf{Optimized Value} \\
\midrule
$a_m$         & 2.322                 & 3.474                    \\
$b_m$         & 0.764                  & 0.764                    \\
$\sigma_m$    & 0.143                  & 0.308                    \\
$a_t$         & -1.061                 & -1.604                    \\
$b_t$         & 0.852                  & 1.200                    \\
$\sigma_t$    & 0.202                  & 0.618                    \\
$b_a$         & 0.926                  & 1.509                    \\
$\sigma_a$    & 0.481                  & 0.010                    \\
$u$           & 0.200                  & 0.395                    \\
$a$           & 0.005                  & 0.62               \\
$d$           & 10.000                 & 29.64                   \\
$s$           & 0.100                  & 1e-15                    \\
$v$           & 0.500                  & 0.900                    \\
$\kappa$          & 0.100                  & 0.041	                    \\
\bottomrule
\end{tabular}
\end{table}

\begin{figure}
    \centering
    \begin{subfigure}[b]{0.45\textwidth}
        \centering
        \includegraphics[width=\linewidth]{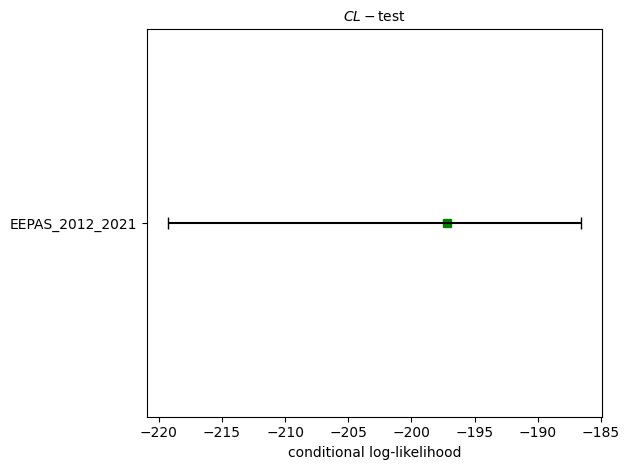}
        \caption{cL-test}
        \label{fig:italy_p_cltest}
    \end{subfigure}
    \hfill 
    \begin{subfigure}[b]{0.45\textwidth}
        \centering
        \includegraphics[width=\linewidth]{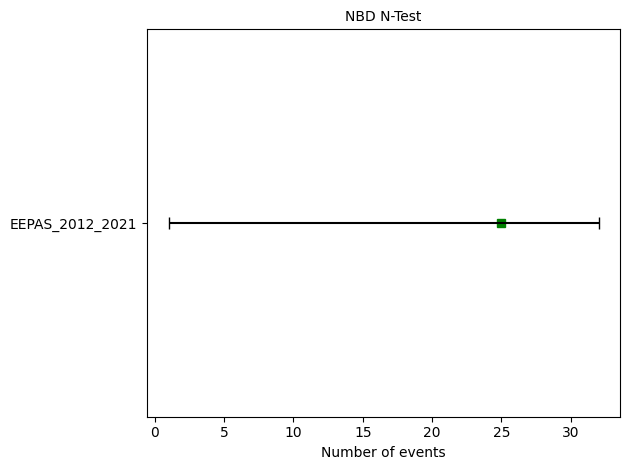}
        \caption{N-test}
        \label{fig:italy_p_ntest}
    \end{subfigure}
   
    \begin{subfigure}[b]{0.45\textwidth}
        \centering
        \includegraphics[width=\linewidth]{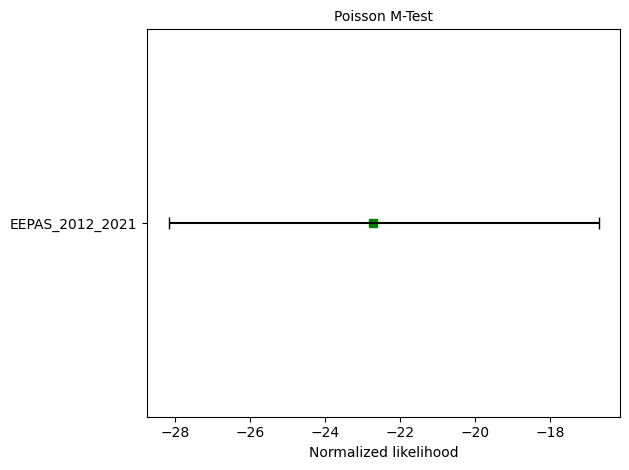}
        \caption{M-test}
        \label{fig:italy_p_mtest}
    \end{subfigure}
    \hfill 
    \begin{subfigure}[b]{0.45\textwidth}
        \centering
        \includegraphics[width=\linewidth]{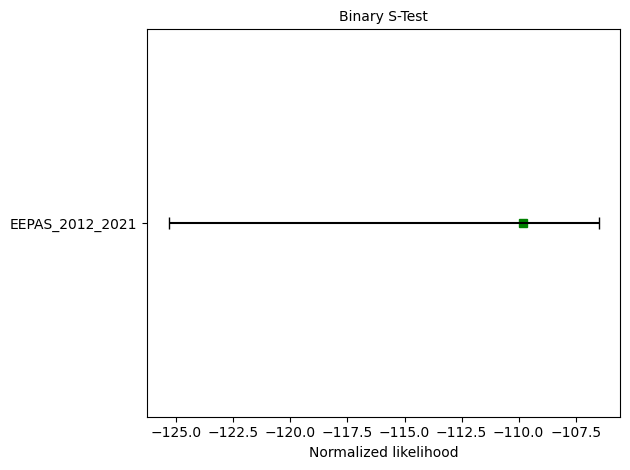}
        \caption{S-test}
        \label{fig:italy_p_stest}
    \end{subfigure}
    \caption{The results of the consistency tests in the testing set (2012–2021) using the full pipeline. 
    (a) Conditional likelihood consistency test (cL-test). 
    (b) Number consistency test (N-test). 
    (c) Magnitude consistency test (M-test). 
    (d) Spatial consistency test (S-test).}
    \label{fig:italy_pipe_tests}
\end{figure}

We then set the prediction intervals to 3 months and generate a 3-month rolling forecast in gridded forecast format. As in the previous section, we apply a set of new CSEP consistency tests based on a binary likelihood function to evaluate forecasts over the testing period. The results are shown in \Cref{fig:italy_pipe_tests}. As observed, the forecasts pass all consistency tests, including the Negative Binomial N-test, binary conditional likelihood (CL) test, and binary spatial (S) test. Since the model assumes a Gutenberg–Richter frequency–magnitude distribution, it also passes the Poisson M-test.

\begin{table}
\centering
 \caption{Performance of PPE and EEPAS models evaluated using scoring rules from \texttt{pyCSEP}.} \label{tbl:score}
\begin{tabular}{rlllll}
\toprule
       & Poisson Likelihood & Binary Likelihood & Kagan Information Score & Brier Score & \\ \hline
PPE    & -168.25            & -110.16           & 1.20                    & -0.011129   & \\ \hline
EEPAS  & -167.90            & -109.80           & 1.33                    & -0.011127   & \\ \hline
\end{tabular}
\end{table}

In \Cref{tbl:score}, we present a quantitative comparison of the PPE and EEPAS models using Poisson and binary joint log-likelihood scores, Kagan information scores, and Brier scores. Binary (spatial) joint log-likelihood values are generally higher than Poisson counterparts because the binary likelihood is less sensitive to clustering effects in seismicity. Nevertheless, both scoring systems indicate that EEPAS outperforms the PPE model. In particular, the higher Kagan information and Brier scores obtained by EEPAS suggest a better agreement between the model and observed data. These scores reflect how much more informative a spatially localized seismicity model is compared to the Spatially Uniform Poisson (SUP) model, which assumes earthquakes are independently and uniformly distributed in space and time.

\section{Conclusion}
This work addresses long-standing gaps in the formulation, implementation, and accessibility of medium to long term earthquake forecasting models, particularly EEPAS and PPE, which, despite their strong empirical performance, remain underdocumented and computationally opaque. We present a mathematically rigorous foundation for these models by deriving analytical expressions for key components such as the incompleteness correction $\Delta(m)$ and normalization factor $\eta(m)$, and by formally situating the EEPAS log-likelihood within the framework of inhomogeneous Poisson point processes. We further clarify the probabilistic basis of precursor scaling by connecting empirical $\Psi$ regressions to likelihood-based inference, and we provide a proof that joint maximum likelihood estimation yields rate normalization under appropriate conditions.

Building on this theoretical groundwork, we introduce the first fully documented and high-performance Python implementation of EEPAS. The framework features JIT acceleration with Numba, modular configuration via JSON, vectorized operations using \texttt{NumPy}, and parallel processing with \texttt{joblib}, and is fully compatible with \texttt{pyCSEP} for standardized testing and comparison. Applied to the Italy HORUS dataset, the system successfully replicates published results within one hour using the same initial parameters. More importantly, it also supports a complete end to end workflow that autonomously estimates model parameters from raw data. Our experiments show that this automated pipeline not only yields better log-likelihood scores but also passes statistical consistency tests, demonstrating the robustness and generalizability of the approach. The initial parameters estimated by the framework can be directly used for downstream forecasting, offering a scalable alternative to the manual identification of $\Psi$ trends in regional catalogs, which has long been a barrier to broader adoption.

We position our framework as part of a growing open source ecosystem for seismological research that spans the full pipeline from data acquisition to forecast evaluation. This ecosystem includes libraries such as \texttt{ObsPy} \cite{beyreuther2010obspy} and \texttt{Pyrocko} \cite{heimann2017pyrocko} for seismic time series analysis, \texttt{SeismoStats} for parameter estimation, \texttt{pyCSEP} for prospective forecast testing, \texttt{SeisBench} \cite{woollam2022seisbench} for machine learning based waveform analysis, and \texttt{OpenQuake} \cite{pagani2014openquake} for seismic hazard and risk modeling. Together, these tools exemplify a modular, transparent, and community driven approach to computational seismology. Our framework fills a key gap in this ecosystem by providing robust tools for statistical modeling of earthquake catalogs, an essential but underserved component in probabilistic seismic forecasting. Released under the MIT license, it offers an accessible, reproducible, and extensible implementation of the EEPAS model, enabling researchers to build end to end, open workflows for seismic hazard analysis and model development. By contributing to this collaborative infrastructure, we aim to advance open science and lower the barrier to entry for reproducible research in earthquake forecasting.


\begin{acknowledgments}
The first author acknowledges support from the National Science and Technology Council, Taiwan, under grant number NSTC 114-2118-M-110 -002 -MY3. The second author is supported by the grant NSTC 114-2115-M-110-003 and the third author is supported by the grant NSTC 114-2116-M-194-011.
\end{acknowledgments}

\begin{dataavailability}
The seismic catalogs employed in this study are publicly available. The HORUS catalog \cite{lolli2020homogenized} is available from \url{https://horus.bo.ingv.it}, and the Italian Parametric Earthquake Catalog (CPTI15, \textit{Catalogo Parametrico dei Terremoti Italiani}) \cite{rovida2020italian} is available from \url{https://emidius.mi.ingv.it/CPTI15-DBMI15/}. The 177 cells and polygon defining the testing and neighborhood region are sourced from \cite{2023ER}. The code developed for this research is provided as supplementary material and will be made publicly available on GitHub upon manuscript acceptance.
\end{dataavailability}

\bibliographystyle{gji}
\bibliography{ref}

\appendix
\section{The PPE and EEPAS model} 
\label{sec:model}
\subsection{EEPAS model}
The EEPAS model is based on empirical linear regression relationships observed in earthquake data. These relationships link the characteristics of precursor earthquakes to the properties of subsequent mainshocks. Evison and Rhoades \cite{ER} studied long-term seismogenesis in four regions and proposed the following predictive relations:
\begin{eqnarray}
\label{am}    M_m = a_M + b_M M_p,\\
\label{at}    \log_{10}(T_p) = a_T + b_T M_p,\\
\label{aa}    \log_{10}(A_p) = a_A + b_A M_p,
\end{eqnarray}
where $M_m$ is the mainshock magnitude, $M_p$ is the mean magnitude of the largest three precursors, $T_p$ is the time interval between the onset of the precursory scale increase and the mainshock, $A_p$ is the spatial extent of the precursors, and $a_M, b_M, a_T, b_T, a_A, b_A$ are regression coefficients obtained by fitting empirical data. Equations \eqref{am}--\eqref{aa} describe the relationships in time, magnitude, and space between major earthquakes and their precursory shocks.

To reduce noise from minor events, earthquakes with magnitudes smaller than $m_0$ are excluded due to instrument precision limitations. Let the $i$-th earthquake (indexed by its occurrence time), observed in a data collection region $S$, occur at time $t_i$, with magnitude $m_i \geq m_0$ and location $(x_i, y_i)$. Each such event contributes an instantaneous increment $\lambda_i(t,m,x,y)$ to the rate density of future seismicity in its neighborhood, defined as
\begin{equation}\label{li}
    \lambda_i(t,m,x,y) = w_i f_i(t) g_i(m) h_i(x,y),
\end{equation}
where $w_i$ denotes a weighting factor determined by the aftershock model proposed in \cite{Rhoades}. Recognizing that real-world data exhibit variability due to the complex and chaotic nature of seismic processes, the EEPAS model generalizes these deterministic relations into probabilistic formulations.

The temporal component $f_i(t)$ is modeled using a log-normal probability density function:
\begin{equation}\label{ft}
    f_i(t) = \frac{H(t - t_i)}{(t - t_i)\ln 10\, \sigma_T \sqrt{2\pi}} \exp{\left[ -\frac{1}{2} \left( \frac{\log_{10}(t - t_i) - a_T - b_T m_i}{\sigma_T} \right)^2 \right]},
\end{equation}
the magnitude component $g_i(m)$ is given by
\begin{equation}\label{gm}
    g_i(m) = \frac{1}{\sigma_M \sqrt{2\pi}} \exp{\left[ -\frac{1}{2} \left( \frac{m - a_M - b_M m_i}{\sigma_M} \right)^2 \right]},
\end{equation}
and the spatial component $h_i(x,y)$ is defined as
\begin{equation}\label{hxy}
    h_i(x,y) = \frac{1}{2\pi \sigma_A^2 10^{b_A m_i}} \exp{\left[ -\frac{(x - x_i)^2 + (y - y_i)^2}{2 \sigma_A^2 10^{b_A m_i}} \right]}.
\end{equation}
Here, $H$ denotes the Heaviside function.

\subsection{PPE model}

The PPE model \cite{JK} describes the long-term background rate of earthquake occurrence and is formulated as
\begin{equation}\label{l0}
    \lambda_0(t,m,x,y) = f_0(t)\, g_0(m)\, h_0(x,y),
\end{equation}
where $f_0(t)$, $g_0(m)$, and $h_0(x,y)$ represent the temporal, magnitude, and spatial components, respectively.  
The temporal function $f_0(t)$ takes the form
\begin{equation}\label{f0}
    f_0(t) = \frac{1}{t - t_0},
\end{equation}
where $t_0$ is the initial time from which the earthquakes are considered.  
The magnitude function $g_0(m)$ is defined as
\begin{equation}\label{g0}
    g_0(m) = \beta \exp\!\left[-\beta(m - m_T)\right],
\end{equation}
where $m_T$ is the target magnitude threshold, and $\beta = b_{\mathrm{GR}}\ln 10$, with $b_{\mathrm{GR}}$ denoting the Gutenberg--Richter $b$-value.  
The spatial component $h_0(x,y)$ is expressed as a summation over past events:
\begin{eqnarray}\label{h0}
 \nonumber h_0(x,y) & = & \sum\nolimits^{\,t-\mathrm{delay}}_{\;t_i \in [t_0,t),\,m_i \ge m_T} h_{0i}(x,y) \\
& \equiv & \sum\nolimits^{\,t-\mathrm{delay}}_{\;t_i \in [t_0,t),\,m_i \ge m_T} \!\left[a(m_i - m_T)\frac{1}{\pi}\!\left(\frac{1}{d^2 + (x - x_i)^2 + (y - y_i)^2}\right) + s\right].
\end{eqnarray}
Here, $d$ is a smoothing parameter that controls the spatial influence of nearby past earthquakes, and $s$ is a constant that ensures a nonzero probability of events occurring far from previous epicenters. $a$ is a normalization factor chosen to satisfy
\begin{equation}\label{nc}
\iint_R h_0(x,y)\,dA
= \frac{n_c(t)}{\displaystyle\int_{t_{(1)}}^{\,t-\mathrm{delay}}\! f_0(u)\,du},
\end{equation}
where $n_c(t)$ is the number of earthquakes with magnitudes exceeding $m_T$ within the target region $R$ during the catalog window $[t_0,\,t-\mathrm{delay})$. The term $t_{(1)} = \min\{\,t_i \in [t_0, t) : m_i \ge m_T\,\}$ denotes the time of the first counted event, used to avoid the singularity at $t_0$. The physical interpretation of the model can be understood by deriving the instantaneous background rate of $M \ge m_T$ events over the entire region $R$, denoted as $\mathrm{Rate}(t)$:
\begin{eqnarray*}
\mathrm{Rate}(t) & \equiv & \int_{m_T}^{m_u}\!\!\iint_R \lambda_0(t,m,x,y)\,dA\,dm  =  f_0(t)\!\left(\int_{m_T}^{m_u}\! g_0(m)\,dm\right)\!\left(\iint_R h_0(x,y)\,dA\right) \\
& \approx & f_0(t)\,\frac{n_c(t)}{\displaystyle\int_{t_{(1)}}^{\,t-\mathrm{delay}}\! f_0(u)\,du},
\end{eqnarray*}
or large $m_u$. Between successive events, $n_c(t)$ remains constant while $f_0(t) = 1/(t - t_0)$ decreases, causing the rate to decay gradually. When a new earthquake with $M \ge m_T$ occurs, $n_c(t)$ increases by one, leading to a discrete upward step in the rate.  
The normalization term $\int_{t_{(1)}}^{t-\mathrm{delay}} f_0(u)\,du$ ensures that integrating $\mathrm{Rate}(u)$ over the catalog window yields $n_c(t)$. 

In summary, $h_0$ encodes long-term spatial memory of past epicenters (with the ${}^{t-\mathrm{delay}}$ exclusion to suppress short-term clustering), $g_0$ distributes magnitudes according to the Gutenberg--Richter law, and $f_0$ accounts for temporal evolution consistent with the observed cumulative counts.

\subsection{Forecasting model allowing for aftershocks}

As mentioned in \cite{RE}, there are two ways to determine the weight parameter $w_i$. One approach is to set $w_i = 1$, meaning that every earthquake contributes equally to the model. The other approach assigns lower weights to earthquakes that are likely aftershocks of previous events. To implement this second strategy, an aftershock model is required. The authors of \cite{RE} adopted the framework from \cite{CM,Ogata1,Ogata2,Ogata3} and proposed the following model:
\begin{equation}\label{l'tmxy}
    \lambda'(t,m,x,y) = \nu \lambda_0(t,m,x,y) + \kappa \sum_{t_i \geq t_0} \lambda_i'(t,m,x,y),
\end{equation}
where $\nu$ denotes the proportion of earthquakes that are not aftershocks, $\lambda_0$ is the baseline rate density given by the PPE model in \eqref{l0}--\eqref{h0} (with the ${}^{t-\mathrm{delay}}$ convention as above), $\kappa$ is a normalization constant, and $\lambda_i'(t,m,x,y)$ is the aftershock rate density contributed by the $i$-th earthquake, defined as
\begin{equation}\label{l'itmxy}
    \lambda_i'(t,m,x,y) = f'_i(t)\, g'_i(m)\, h'_i(x,y).
\end{equation}
Here, $f'_i(t)$ is the temporal density function given by the modified Omori law:
\begin{equation}\label{f2it}
    f'_i(t) = H(t - t_i)\, \frac{p - 1}{(t - t_i + c)^p},
\end{equation}
where $t_i$ is the origin time of the $i$-th earthquake, and $c, p$ are Omori law parameters.  

The magnitude component $g'_i(m)$ is defined as
\begin{equation}\label{g2im}
    g'_i(m) = H(m_i - \delta - m)\, \beta \exp\left[ -\beta (m - m_i) \right],
\end{equation}
where $m_i$ is the magnitude of the $i$-th earthquake, and $\delta$ is a threshold parameter based on Bath’s law, determining whether a subsequent event can be classified as an aftershock.  

The spatial component $h'_i(x,y)$ is assumed to follow a bivariate normal distribution:
\begin{equation}\label{h2ixy}
    h'_i(x,y) = \frac{1}{2\pi \sigma_U^2 10^{m_i}} \exp\left[ -\frac{(x - x_i)^2 + (y - y_i)^2}{2 \sigma_U^2 10^{m_i}} \right],
\end{equation}
where $\sigma_U$ is a parameter calibrated to satisfy Utsu’s regional relation.

Using the aftershock model above, the weight $w_i$ assigned to the $i$-th earthquake is defined by
\begin{equation}\label{wi}
    w_i = \frac{\nu\, \lambda_0(t_i, m_i, x_i, y_i)}{\lambda'(t_i, m_i, x_i, y_i)},
\end{equation}
implying that an earthquake receives a weight close to zero if it is likely an aftershock, and a weight close to one if it appears not to be an aftershock.

\label{lastpage}

\end{document}